\documentclass[twocolumn,showkeys,showpacs,preprintnumbers,aps,prd,superscriptaddress,nofootinbib]{revtex4-1}
\usepackage{graphicx}
\usepackage{epsf}
\usepackage{bm}
\usepackage{amsmath}
\usepackage{amsfonts}
\usepackage{amssymb}
\usepackage{epstopdf}
\usepackage{hyperref}
\usepackage{color}
\usepackage{verbatim}
\usepackage{multirow}

\def\nar{\ref@jnl{New A Rev.}}          
\def\aap{\ref@jnl{A\&A}}                
\def\araa{\ref@jnl{ARA\&A}}             

\begin{document}

\title{Probing the Cosmic Distance Duality Relation via Non-Parametric Reconstruction for High Redshifts}


\author{Felipe Avila}
\email{felipeavila@on.br}
\affiliation{Observat\'orio Nacional, Rua General Jos\'e Cristino 77, 
S\~ao Crist\'ov\~ao, 20921-400 Rio de Janeiro, RJ, Brazil}

\author{Fernanda Oliveira}
\email{fernandaoliveira@on.br}
\affiliation{Observat\'orio Nacional, Rua General Jos\'e Cristino 77, 
S\~ao Crist\'ov\~ao, 20921-400 Rio de Janeiro, RJ, Brazil}

\author{Camila Franco}
\email{camilafranco@on.br}
\affiliation{Observat\'orio Nacional, Rua General Jos\'e Cristino 77, 
S\~ao Crist\'ov\~ao, 20921-400 Rio de Janeiro, RJ, Brazil}

\author{Maria Lopes}
\email{marialopes@on.br}
\affiliation{Observat\'orio Nacional, Rua General Jos\'e Cristino 77, 
S\~ao Crist\'ov\~ao, 20921-400 Rio de Janeiro, RJ, Brazil}

\author{Rodrigo Holanda}
\email{holandarfl@gmail.com}
\affiliation{Universidade Federal do Rio Grande do Norte, Departamento de F\'{i}sica Te\'{o}rica e Experimental, 59300-000, Natal - RN, Brazil}
\affiliation{Departamento de F\'{\i}sica, Universidade Federal de Campina Grande, 58429-900, Campina Grande - PB, Brasil}

\author{Rafael C. Nunes}
\email{rafadcnunes@gmail.com}
\affiliation{Instituto de F\'{i}sica, Universidade Federal do Rio Grande do Sul, 91501-970 Porto Alegre RS, Brazil}
\affiliation{Divis\~{a}o de Astrof\'{i}sica, Instituto Nacional de Pesquisas Espaciais, Avenida dos Astronautas 1758, S\~{a}o Jos\'{e} dos Campos, 12227-010, S\~{a}o Paulo, Brazil}

\author{Armando Bernui}
\email{bernui@on.br}
\affiliation{Observat\'orio Nacional, Rua General Jos\'e Cristino 77, 
S\~ao Crist\'ov\~ao, 20921-400 Rio de Janeiro, RJ, Brazil}

\begin{abstract}
We test the validity of the cosmic distance duality relation (CDDR) by combining angular diameter distance and luminosity distance measurements from recent cosmological observations. For the angular diameter distance, we use data from transverse baryon acoustic oscillations and galaxy clusters. 
	On the other hand, the luminosity distance is obtained from Type Ia supernovae in the Pantheon+ sample and from quasar catalogs. 
	To reduce the large dispersion in quasar luminosity distances, we apply a selection criterion based on their deviation from the $\Lambda$CDM model and implement a binning procedure to suppress statistical noise. We reconstruct the CDDR using Gaussian Processes, a non-parametric supervised machine learning method. Our results show no significant deviation from the CDDR within the $2\sigma$ confidence level across the redshift range explored, supporting its validity even at high redshifts.
\end{abstract}

\keywords{distance duality; gaussian processes; binning data}

\pacs{}

\maketitle

\section{Introduction}
\label{sec1:introducao}

One of the most pivotal achievements in modern cosmology is the precise determination of distances to remote astronomical sources. By comparing a source's redshift with its measured distance—obtained either from standard candles (luminosity distance, $D_L$) or standard rulers (angular diameter distance, $D_A$)—we can place stringent constraints on cosmological models. For example, the existence of dark energy and dark matter has been confirmed with remarkable precision through observations of Type Ia supernovae, which serve as standard candles, and baryon acoustic oscillations, which act as standard rulers (see, e.g.,~\cite{DiValentino:2025sru,2009ARNPS..59..397C,2013PhR...530...87W}). These discoveries highlight the crucial role of distance measurements from diverse cosmic probes in advancing our understanding of the universe, as they provide insights into the nature of roughly 95\% of its matter-energy content. Despite the success of the standard flat-$\Lambda$CDM model in accounting for a wide array of observations, the increasing precision of modern datasets has brought to light several persistent tensions—such as discrepancies in the Hubble constant and the amplitude of structure growth—that challenge the model's completeness and may point toward new physics beyond the current framework~\cite{DiValentino:2025sru,2022NewAR..9501659P,Vagnozzi:2023nrq}.

The luminosity distance function, $D_L(z)$, and the angular diameter distance function, $D_A(z)$, are related by the so-called cosmic distance duality relation (CDDR), given by $D_L(z) = D_A(z)(1+z)^2$~\citep{Etherington33,Ellis07}. This relation holds in all cosmological models based on Riemannian geometry and does not rely on Einstein's field equations or on the specific properties of the universe’s matter-energy content. For the CDDR to remain valid, three fundamental conditions must be satisfied~\citep{Qin21,Bassett04}:  
(i)~spacetime must be described by a metric theory of gravity;  
(ii)~photons must travel along null geodesics; and  
(iii)~the number of photons must be conserved.  
This consistency condition underpins the interpretation of key cosmological probes, including the cosmic microwave background, galaxy clusters, and gravitational lensing. 
Consequently, any observed deviations from the CDDR could signal the presence of new physics beyond the standard model or point to unaccounted-for systematic effects in the data\footnote{For discussions on violations of the CDDR in modified gravity scenarios, see, e.g.,~\cite{2014PhRvD..90l4064H,2021PhRvD.104h4079A,2020JCAP...11..047L,2016PhRvD..94b3524H,2017CQGra..34s5003H}.}. 
In fact, in the context of Type Ia supernova observations, some possible systematics such as the attenuation of photons due to cosmic opacity may arise from 
dust~\cite{2022JCAP...01..053K,2018PhRvD..97b3538H,2004NewAR..48..583C,2013JCAP...04..042S,2017GReGr..49..150J,2013PhRvD..87j3013L,2013PhLB..718.1166L}.

On the other hand, 
several theoretical and observational efforts have been made to investigate possible violations of the CDDR \cite{Rasanen:2015kca,Renzi:2021xii,Qi:2024acx,Keil:2025ysb,Rana:2015feb,Gahlaut:2025lhv,Mukherjee:2021kcu,More:2016fca,Holanda10}.
These tests are generally categorized into two classes: those that assume a specific cosmological model and those that do not. Model-dependent analyses, often based on the $\Lambda$CDM framework, can place tighter constraints on deviations of the CDDR, but their conclusions are limited by the assumed cosmological scenario \cite{2010JCAP...10..024A,Bassett04,2016JCAP...02..054H,2011A&A...528L..14H,2016APh....84...78H}. Model-independent methods, by contrast, rely directly on astrophysical measurements without invoking an underlying model, though they tend to yield results with larger statistical uncertainties \cite{Holanda10,2011ApJ...729L..14L,2012JCAP...06..022H,2013PhRvD..87j3530E,2015PhRvD..92b3520W,2019APh...108...57Y,2020EPJC...80..838X,2020PhRvD.102f3513D}. 
Additionally, most of these analyses are restricted to the redshift interval covered by type Ia supernovae. 
Although current observations are broadly consistent with the expected relation, expanding these tests to include alternative observables and higher redshifts is crucial. 
Such efforts can help detect subtle deviations, confirm the robustness of the cosmological framework, or reveal hidden sources of systematic errors. 
%

More recently, several non-parametric methods have been employed to test the validity of the CDDR~\cite{2015JCAP...10..061S,2021EPJC...81..903L}. In~\cite{Teixeira:2025czm}, the authors investigate how potential violations of the distance duality relation could help reconcile the Hubble tension and affect the inferred properties of dark energy from cosmological observations. Similarly, ref.~\cite{2012IJMPD..2150008H} examined how violations of the CDDR impact estimates of the Hubble constant, $H_0$. Along the same lines, ref.~\cite{2025arXiv250416868A} search for systematics in DESI DR2 and Pantheon+ data by applying a distance duality consistency test to ensure robust inferences about dark energy. These studies highlight the importance of further probing the CDDR, as it offers valuable insights into multiple aspects of observational cosmology.


In this work, we use the following observables for the angular diameter distance data: 
transverse BAO and Galaxy clusters (GC). 
For the luminosity distance data, we consider standard candles—specifically, 
type Ia supernovae (SNIa) and quasars (QSO). 
To incorporate quasar-based $D_L(z)$ measurements into the CDDR analysis, we adopt two complementary approaches: (i) we apply a selection criterion based on the dispersion of SNIa $D_L(z)$ values relative to the $\Lambda$CDM model, and (ii) we implement a data binning technique that significantly reduces the intrinsic scatter by compressing a large number of measurements into a smaller set of representative points. 
As emphasized by \cite{Bassett04}, potential deviations from the CDDR are expected to be more prominent at high redshifts, making the inclusion of QSO in the $D_L(z)$ sample particularly crucial.

This paper is organized as follows. 
In Section~\ref{sec2:data}, we describe the dataset used to test the CDDR. 
Section \ref{sec3:binning_data} outlines the binning strategy employed to effectively incorporate QSO measurements. 
In Section \ref{sec4:GP}, we briefly introduce the non-parametric reconstruction method known as Gaussian Processes (GP), which is used to reconstruct the duality relation. Section \ref{sec5:results} presents the main reconstruction results, followed by a discussion of the findings in Section \ref{sec6:conclusions}. 

\section{Data Sets}
\label{sec2:data}

In this section, we present the observational data sets used in this work, namely the angular diameter distances, $D_A(z)$, and luminosity distances, $D_L(z)$. 
Along with a brief description of each data set, we also describe the methodology of our data analyses. 
This approach is motivated by the goal of optimizing the performance of the binning procedure applied in our reconstruction.
In Table~\ref{tab:data_summary}, we summarize the data samples and analyses performed in this work.


\subsection{Transverse Baryon Acoustic Oscillations data}
\label{subsubsec:BAO_data}

The BAO data, $\theta_{\rm BAO}$, correspond to the angular scale of 
transverse BAO scale, a measurement derived from the peak observed in the angular correlation function, $\omega(\theta)$~\cite{Edilson18,DESI2}. Transverse BAO measurements are obtained by applying a parametric fit proposed by \cite{Sanchez11} to the angular correlation function. This methodology is able to extract the BAO peak with weak cosmological dependence. 
This observable is related to the angular diameter distance 
\begin{equation}\label{eq:da_bao}
	\frac{D_{A}(z)}{r_{d}} = \left[(1+z)\,\theta_{\rm BAO}(z)\right]^{-1}\,,
\end{equation}
where $r_d$ is the sound horizon at the drag epoch. By propagating the uncertainty in $\theta_{\rm BAO}$, denoted $\delta\theta_{\rm BAO}$, we obtain
\begin{equation}\label{eq:Erro_da_bao}
	\delta\left[\frac{D_{A}(z)}{r_{d}}\right](z) 
	= \frac{D_{A}(z)}{r_{d}}\,\theta_{\rm BAO}^{-1}(z)\,\delta\theta_{\rm BAO}(z)\,.
\end{equation}

Our analysis uses 16 measurements of $\theta_{\rm BAO}$: 14 data points from~\cite{Menote22} in the redshift range $0.35 \leq z \leq 0.63$, one data point from~\cite{Edilson18} at $z = 2.225$, and one from~\cite{Edilson21} at $z = 0.11$. 
In addition, we complement the BAO sample with 6 measurements of $D_M/r_d$ from DESI DR2\footnote{Note that $D_M(z) \equiv (1+z)\,D_A(z)$.}~\citep{DESI2}, spanning the redshift range $0.510 \leq z \leq 2.33$. Throughout this work, we adopt the fiducial value $r_d = 147.05$ Mpc~\citep{Planck20}. It is important to note that only pre-recombination physics is sensitive to changes in the sound horizon scale $r_d$. All datasets used in this work to test the CDDR are of the late-time type, meaning they are independent of physics that could alter $r_d$. Moreover, as shown in \citep{Bonilla:2020wbn}, varying $r_d$ within reasonable ranges has no significant impact on parametric reconstructions like those performed here.





\subsection{Galaxy Clusters data}
\label{subsubsec:cluster_data}

The Sunyaev–Zel'dovich effect (SZE) is a small distortion in the cosmic microwave background (CMB) spectrum, caused by inverse Compton scattering of CMB photons from high-energy electrons, typically found in the intracluster medium of galaxy clusters~\cite{1972CoASP...4..173S}. 
This interaction between CMB photons and the hot intracluster medium leaves a characteristic spectral distortion in the CMB proportional to the integrated Compton-$y$ parameter along the line of sight~\cite{1999PhR...310...97B}

\begin{equation}
    \frac{\Delta T_{\rm SZE}(\hat{n})}{T_0} = f_\nu y(\hat{n}) = f_\nu \int \frac{k_B T_e}{m_e c^2} \sigma_T n_e \, dl \,,
\end{equation}
where:
\begin{itemize}
    \item $\Delta T_{\rm SZE}/T_0$ is the fractional temperature change of the CMB
    \item $T_0 = 2.7255$ K is the present-day CMB temperature
    \item $f_\nu$ describes the spectral dependence of the effect (negative at $\nu < 218$ GHz, positive at higher frequencies)
    \item $y(\hat{n})$ is the Comptonization parameter along direction $\hat{n}$
    \item $k_B$ is Boltzmann's constant, $m_e$ the electron mass, $c$ the speed of light
    \item $\sigma_T$ is the Thomson cross-section
    \item $n_e$ and $T_e$ are the electron number density (cm$^{-3}$) and temperature (K) of the intracluster plasma
    \item The integral follows the line of sight $dl$ through the cluster
\end{itemize}

The amplitude of this effect is typically $\Delta T/T_0 \sim \mathcal{O}(10^{-5})$ for massive clusters, with several notable features, like (i) the thermal SZE is redshift-independent, making it particularly valuable for detecting high-z clusters, (ii) the effect scales with the integrated pressure $P_e = n_e k_B T_e$ of the intracluster medium, (iii) The spectral signature allows separation from primary CMB anisotropies through multi-frequency observations.

In contrast to the SZE, the X-ray emission from the same intracluster medium (ICM) originates primarily through two distinct thermal processes: thermal bremsstrahlung (free-free emission) and line emission from highly ionized metals. The X-ray surface brightness $S_X$ depends quadratically on the electron number density and is given by~\cite{1988xrec.book.....S}
\begin{equation}
    S_X(\hat{n}, E) = \frac{\theta^2 D_A^2}{D_L^24\pi} \int n_e^2(r) \Lambda_{eH}(T_e,Z,E) \, dl \,,
\end{equation}
where:
\begin{itemize}
    \item $\theta$ is the angular size of the cluster 
    \item $S_X(\hat{n}, E)$ is the X-ray surface brightness in direction $\hat{n}$ at energy $E$ (typically 0.5-10 keV for clusters)
    \item $z$ is the cluster redshift (accounting for cosmological dimming)
    \item $n_e(r)$ is the radially-dependent electron number density (cm$^{-3}$)
    \item $\Lambda_{eH}(T_e,Z,E)$ is the X-ray cooling function (erg cm$^3$ s$^{-1}$), which depends on:
    \begin{itemize}
        \item Electron temperature $T_e$ (typically $10^7$-$10^8$ K for clusters)
        \item Metallicity $Z$ (primarily Fe, O, and Si lines)
        \item Energy band $E$ (continuum vs. line emission)
        \end{itemize}
\end{itemize}

The integral above follows the line of sight $dl$ through the cluster. As it is largely known, the different dependencies on $n_e$ in the SZE and X-ray signals enable the inference of the observational angular diameter distance  to galaxy clusters~\cite{2002ApJ...581...53R,2002ARA&A..40..643C}. However, from Eq.(4), as one may see, if the CDDR is not valid (such as 
\(\eta(z) = D_L (1+z)^{-2} / D_A\)), then what is actually obtained from observations is a more general expression for the angular diameter distance \(D_A(z)\), which explicitly must  incorporate the possible violation of the relation. Thus, in any test of the CDDR that employs \(D_A(z)\) obtained via SZE plus X-ray observations from  galaxy clusters, it is necessary to include the factor \(\eta(z)\) in the formalism. Therefore, the more general expression for \(D_A(z)\) is (see details in \cite{2004PhRvD..70h3533U}) 
\begin{equation}
\label{DA_1}
	D_A(z) \propto \frac{(\Delta T_0)^2 \Lambda_{eH,0}}{\theta_c \ S_{X,0} T_{e,0}^2 \eta(z)^2} \, (1+z)^{-4} \, .
\end{equation}
The Equation~(\ref{DA_1}) enables the determination of the observational angular diameter distance to the galaxy cluster if $\eta(z)=1$, specifically for the sample selected under the conditions discussed above.

This technique enables the construction of a Hubble diagram based on galaxy clusters, allowing the estimation of cosmological parameters such as the matter density, dark energy density, and the Hubble constant $H_0$~\cite{2013JCAP...06..033H,2012JCAP...02..035H,colaço2023hubble}. Importantly, it provides a standard ruler-based distance measurement that is independent of standard candles and can be applied at high redshifts~\cite{1999PhR...310...97B,2013JCAP...06..033H,2012JCAP...02..035H}.

However, extracting angular diameter distances from SZE/X-ray observations requires assumptions about the geometry and morphology of the intracluster gas distribution. In our analysis, we consider a set of angular diameter distance measurements from a sample of galaxy clusters compiled by~\cite{2005ApJ...625..108D}, which we use for Gaussian Process reconstruction.

This sample includes 25 galaxy clusters spanning the redshift range $0.023 \leq z \leq 0.784$. To model the physical properties of the clusters and extract the angular diameter distances, an isothermal elliptical $\beta$-model is adopted. In particular, ref.~\cite{2005ApJ...625..108D} provides angular diameter distances obtained from three different approaches:  
(i) assuming the standard $\Lambda$CDM cosmology (second column of Table III in~\cite{2005ApJ...625..108D});  
(ii) applying the SZE/X-ray method under the assumption of spherical symmetry (third column); and  
(iii) using the SZE/X-ray method with an ellipsoidal (oblate spheroidal) model for the cluster geometry (fourth column).  

In this work, we employ the angular diameter distances from the third approach, where clusters are modeled as oblate spheroids. The estimates for systematic effects  to galaxy cluster data are~\citep{Bonamente06}: SZ calibration $\pm 8\%$, X-ray flux calibration $\pm 5\%$, radio halos $+3\%$, and X-ray temperature calibration $\pm 7.5\%$. Indeed, one may show that typical statistical errors amount to nearly $20\%$, while for systematics we also find typical errors around $+12.4\%$ and $-12\%$. In the present analysis we have combined the statistical and systematic errors in quadrature for the galaxy clusters $\sigma_{\text{data}}^2 = \sigma_{\text{stat}}^2 + \sigma_{\text{syst}}^2$.

As commented earlier, the method based on SZE combined with X-ray surface brightness does not yield the true angular diameter distance $D_A(z)$ directly, but rather an effective quantity defined as (see, e.g.,~\cite{2011A&A...528L..14H,2012A&A...538A.131H}) 
\begin{equation}\label{eq:eta_gc}
	D_A^{\text{cluster}}(z) = \eta^2(z) D_A(z)\,,
\end{equation}
where $D_A^{\text{cluster}}(z)$ refers the quantity obtained when one assumes $\eta=1$ in Eq. (5). Therefore, to test the validity of the CDDR using SZE + X-ray observations, one must replace $D_A(z)$ with $D_A^{\text{cluster}}(z)\,\eta^{-2}(z)$ if the CDDR is written in the form $\eta(z) = D_L(z) / \left[ (1+z)^2 D_A(z) \right]$. 
This leads to the expression 
\begin{equation}\label{eq:eta_da_modified}
	\eta(z) = \frac{D_A^{\text{cluster}}(z)\,(1 + z)^2}{D_L(z)}\,,
\end{equation}
which allows us to evaluate the CDDR using directly galaxy cluster angular diameter distances. In other words, this method aims to avoid circularity in the analysis.

Naturally, if an independent and model-agnostic measurement of $D_A(z)$ for each galaxy cluster from a sample were available, one could use the equation~\eqref{eq:eta_gc} to test functional forms of $\eta(z)$. Therefore, we use galaxy clusters and BAO measurements in the analysis based on equation~\eqref{eq:eta_gc}, and galaxy clusters, SNe Ia and QSO data in the analysis of the expression for $\eta(z)$ given above (Eq.7).


The BAO and GC samples described above constitute our complete set of angular diameter distance measurements.


\begin{figure}[h]
	\centering
	\includegraphics[scale=0.47]{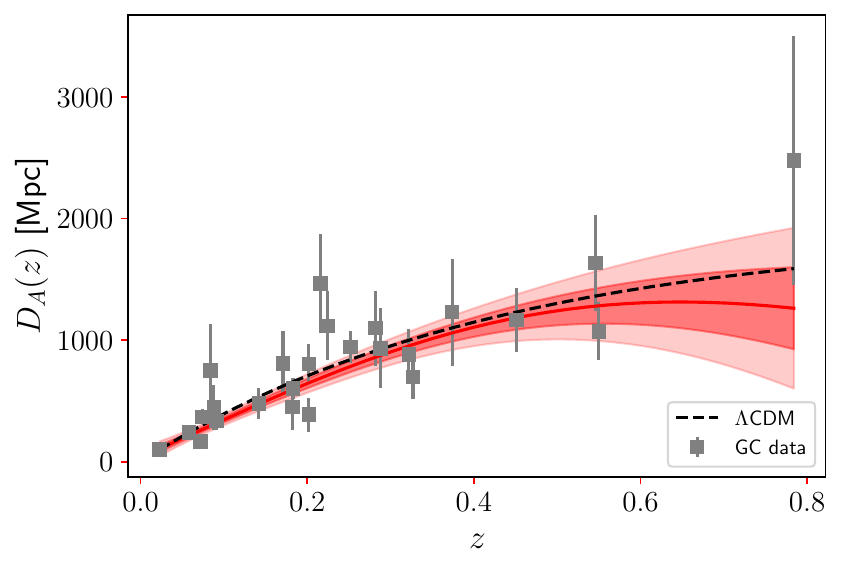} \,\,\,\,\,
	\includegraphics[scale=0.47]{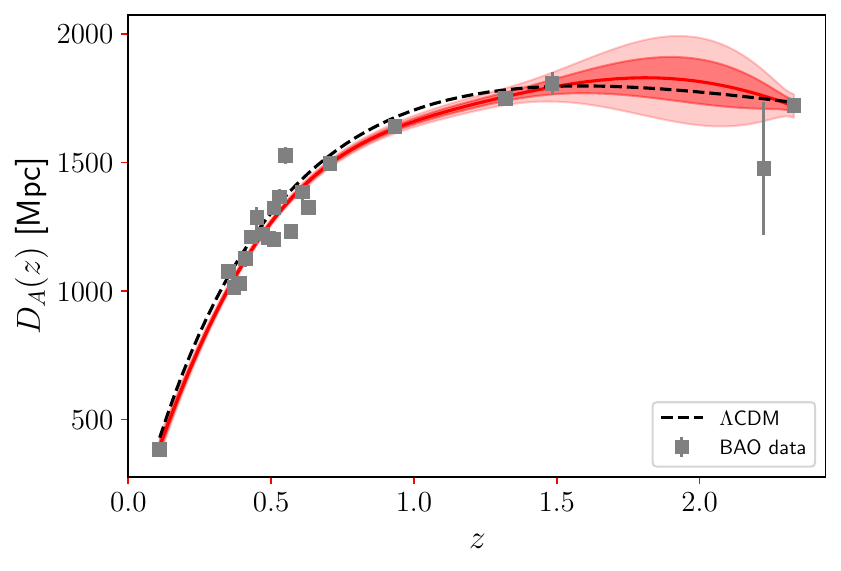}
	\caption{Upper Panel: Reconstruction of the angular diameter distance $D_A(z)$ using our GC sample. The solid red curve represents the mean reconstruction, while the light and dark shaded regions correspond to the $1\sigma$ and $2\sigma$ uncertainties, respectively. The dashed line shows the $\Lambda$CDM prediction for comparison. Lower Panel: Same as in the upper panel, but for the BAO samples.}
	\label{fig:DA_recon}
\end{figure}

\begin{figure}[h]
	\centering
	\includegraphics[scale=0.47]{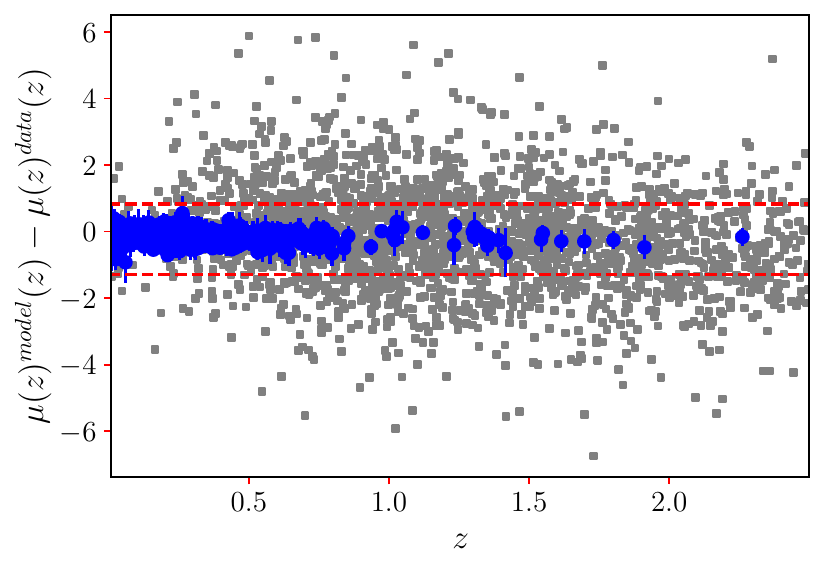} \,\,\,\,\,
	\includegraphics[scale=0.47]{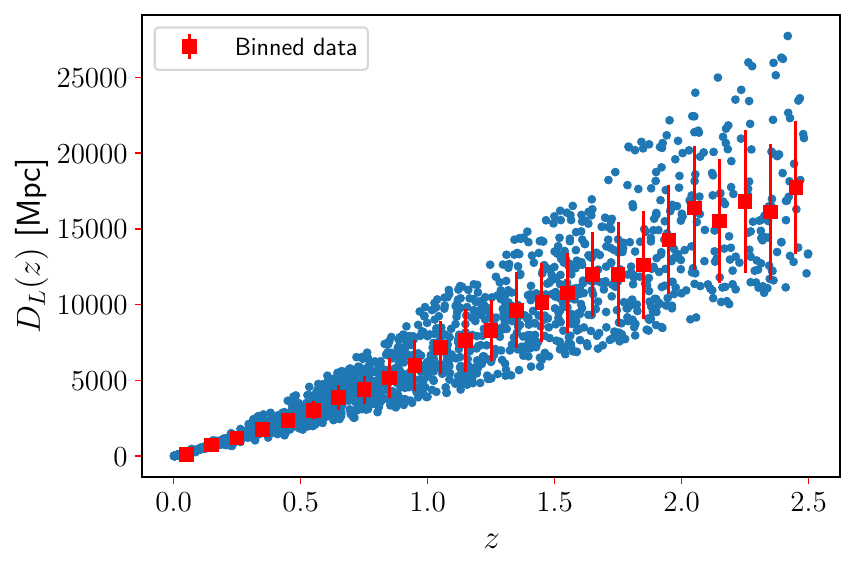}
	\caption{Upper panel: Residuals of the luminosity distance measurements relative to the $\Lambda$CDM predictions for quasars (gray squares) and SNIa (blue dots). The red dashed lines represent the $1\sigma$ dispersion of the SNIa distribution. Only quasars falling within this $1\sigma$ region were retained for further analysis. Lower panel: Binned luminosity distance measurements (red squares) for the final selected dataset, comprising 1701 SNIa light curves and 1160 quasars, totaling 2861 data points.}
	\label{fig:mu_data_cut_DL_bin}
\end{figure}

\subsection{Type Ia Supernovae data}
\label{subsubsec:Supernova_data}

For the Type Ia supernovae (SNIa) data, we adopt the Pantheon+ sample \cite{Pantheon2022}, which comprises 1701 measurements from light curves and spectra collected across multiple surveys, covering the redshift range $z \in [0.001, 2.26]$. This catalogue provides the distance modulus, $\mu(z)$, defined as the difference between the apparent and absolute magnitudes of each supernova 
\begin{equation}\label{eq:mu}
	\mu \equiv m - M = 5 \log_{10} 
	\left( \frac{D_L(z)}{1~ \mathrm{Mpc}} \right) + 25 \,,
\end{equation}
where $D_L(z)$ is the luminosity distance.

\subsection{Quasars data}
\label{subsubsec:quasar_data}

A new class of standard candles is being investigated to significantly extend the Hubble diagram to higher redshifts: quasars (QSO)~\citep{Lusso15,Lusso19,Lusso20,Raffai25,Li25}. 
The methodology relies on the observed non-linear correlation between the X-ray and ultraviolet luminosities of QSO, described by the following relation 
\begin{equation}\label{eq:NL_qso}
	\log(L_X) = \beta + \alpha\log(L_{UV}) \,,
\end{equation}
where $\beta$ and $\alpha$ are free parameters. 
Observations indicate that $\alpha$ is constant, $\alpha \sim 0.6$, leaving 
$\beta$ to change with redshifts, being proportional to $\log(D_L)$. 

Although these distance measurements still pose a significant challenge for cosmological studies due to their large intrinsic dispersion~\citep{Lusso15}, they hold great potential because QSO can be observed at much higher redshifts than SNIa. In this work, we used a recent compilation of 2421 QSO in the range $0.009 \leq z \leq 7.5413$~\citep{Lusso20}. 
Since the $D_A$ measurements of our sample go up to $z=2.33$, we decided to limit the QSO data to $z=2.5$, therefore our QSO sample for analyses has 2195 quasars. 

The SNIa and QSO samples described above constitute our complete set of luminosity distance measurements.

\section{Methodology for data analysis}

In what follows we describe the methodology that will be used to analyze 
the data described above to probe the CDDR via non-parametric reconstruction for high redshift data. 

\subsection{Data Binning Methodology}
\label{sec3:binning_data}

Data binning procedure consists of representing a set of data by a single data point, where its value and error is the mean and the standard deviation of that distribution, respectively. 
Let $(x_i, y_i)$ be a set of data points for $i=1, 2, ..., N$. 
For a given interval $k$ with $N_k$ data points, we can calculate the mean 
\begin{equation}\label{eq:mean_bin}
	\bar{Y}(\bar{x}) = \frac{1}{N_k}\sum_k y_k(x_k) \,,
\end{equation}
and the standard deviation
\begin{equation}\label{eq:std_bin}
	\sigma_{\bar{Y}}(\bar{x}) = \sqrt{\frac{1}{N_k}\sum_k(y_k-\bar{Y})^{2}} \,,
\end{equation}
where $\bar{x}$ is the mean of the interval.

In general, the binning method requires a suitable choice of range to ensure that the data inside each bin approximates a Gaussian distribution. 
This condition is met by both the SNIa and QSO samples, which motivated our use of this method—especially considering that quasars have a high number density, 
exhibiting significant dispersion in the distance modulus.

\begin{table}[ht]
\centering
\caption{Summary of redshift coverage and number of data points for each cosmological probe.}
\label{tab:data_summary}
\begin{tabular}{lcc}
\hline
\textbf{Probe} & \textbf{Redshift Coverage} & \textbf{Number of Data Points} \\
\hline
GC  & $z \in [0.023, 0.784 ]$ & $25$ \\
BAO & $z \in [0.11, 2.33]$ & $22$ \\
SNe & $z \in [0.001, 2.26 ]$ & $1701$ \\
QSO & $z \in [0.009, 2.5]$ & $2195$ \\
\hline
\end{tabular}
\end{table}

\subsection{Gaussian Processes for Cosmological Parameter Reconstruction}
\label{sec4:GP}

Gaussian Processes (GPs) represent a powerful and flexible tool in modern cosmology, providing a non-parametric Bayesian approach to reconstructing functions from observational data while rigorously quantifying uncertainties~\citep{Seikel2012, Seikel13, Yang15, Valente18}. Formally, a GP defines a distribution over functions where any finite set of function values follows a multivariate Gaussian distribution, completely specified by a mean function $m(\mathbf{x})$ and covariance kernel $k(\mathbf{x},\mathbf{x}')$. The primary advantage of GPs lies in their ability to extract maximal information from data without relying on specific parametric cosmological models, while naturally incorporating measurement uncertainties through the likelihood function. This is particularly useful in cosmological studies, where numerous parameters must be inferred from complex, often sparse, observational data. Gaussian Processes have been widely applied to reconstruct and study various cosmological quantities, such as the evolution of the dark energy equation of state, $w(z)$~\cite{Seikel2012, Zhang18, Perenon:2022fgw, Bonilla2022, Abedin:2025yru}, the deceleration parameter, $q(z)$~\citep{Jesus2019, Mukherjee2020}, the growth function $[f\sigma_8](z)$~\citep{Perenon21, Avila22b, Calderon2023msm, LHuillier2019imn, Oliveira25}, the homogeneity scale $R_H(z)$~\citep{Avila22a}, and the time evolution of the growth index, $\gamma(z)$~\citep{Yin19, Avila22b, Mu2023, Oliveira23, Escamilla:2025imi}. Additionally, GPs have been applied to a variety of other problems in cosmology, such as modeling the Hubble parameter and performing model-independent studies of the universe~\citep{Sabogal:2024qxs,Gao:2025ozb,Jiang:2024xnu,Dinda:2024ktd,Dinda:2025svh,Yang:2025kgc, Dinda2023, RuizZapatero2022, Escamilla2023modelindependent, Sun2021, OColgain2021, Kjerrgren2021, Renzi2020, Calderon2022cfj, Dinda2023xqx,Escamilla:2024ahl,Velazquez:2024aya,Gomez-Vargas:2021zyl,Keeley:2020aym,Zheng:2025cgq,Li:2025htp,Yang:2025qdg,Cosmai:2013iga,Wang:2024rxm}. 

In this study, we apply a supervised learning regression approach using GP to reconstruct the distance duality relation, $\eta(z)$, in a non-parametric manner. The GP framework allows us to model and interpolate the observations while also providing an estimate of the uncertainty through empirical confidence intervals. This approach is highly advantageous in cosmological studies, where minimal assumptions are made about the underlying cosmological model, allowing for greater flexibility and a more robust analysis. By relying on a few key physical principles and minimal cosmological assumptions, GP facilitates a data-driven exploration of cosmological parameters.

At the core of a GP lies the concept of a function $f(x)$, which is completely specified by its mean function, $m(\textbf{x})$, and its covariance function, $k(\textbf{x}, \textbf{x}')$. The mean function represents the expected value of the function at any given point $\textbf{x}$, while the covariance function encapsulates the relationship between the values of the function at different points $\textbf{x}$ and $\textbf{x}'$. Mathematically, these are expressed as 
\begin{eqnarray}
	m(\textbf{x}) &=& \mathbb{E}[f(\textbf{x})] \,, \nonumber \\
	k(\textbf{x}, \textbf{x}') &=& \mathbb{E}[(f(\textbf{x}) - m(\textbf{x}))(f(\textbf{x}') - m(\textbf{x}'))] \,.
\end{eqnarray}
Thus, the GP is defined as 
\begin{equation}
	f(\textbf{x}) \sim \mathcal{GP}(m(\textbf{x}), k(\textbf{x}, \textbf{x}')) \,.
\end{equation}

Although GP do not rely on any specific cosmological model, they require a kernel function to define the covariance structure of the data. The choice of kernel significantly influences the performance and accuracy of the GP reconstruction, and various kernels can be explored to assess their impact on the results. Recent studies have examined the role of different kernels in cosmological parameter reconstructions~\citep{Hwang23, Zhang23}. 
One of the kernels most commonly used is the Squared Exponential (SE) kernel, which is highly flexible and widely employed in many cosmological applications due to its smoothness and differentiability properties. The SE kernel is defined as 
\begin{equation}
	k(x, x') \equiv \sigma_f^2 \exp\left( -\frac{(x - x')^2}{2\,l^2} \right)\,,
\end{equation}
where $\sigma_f$ and $l$ are hyperparameters that control the amplitude and length scale of the covariance, respectively. These hyperparameters are optimized during the reconstruction process. The primary advantage of the SE kernel is that it is infinitely differentiable, which is essential for cosmological studies that involve derivatives of quantities like the luminosity distance, $D_L(z)$, and the angular diameter distance, $D_A(z)$. For this reason, the SE kernel is particularly well-suited for our analyses. Given the substantial uncertainties associated with reconstructions, we find that different kernel choices consistently yield compatible results. For a detailed discussion on this matter, see \cite{Bonilla:2020wbn}. Thus, for the data and methodology of this work, variations in the kernel selection prove to be fully equivalent.

To perform the GP regression, we use the GaPP code developed in~\cite{Seikel2012}\footnote{\url{https://github.com/JCGoran/GaPP}}, which implements the GP algorithm from~\cite{Rasmussen06}. This code is robust and efficient, enabling the computation of GP derivatives, which are crucial for our work. By utilizing this code, we can reconstruct the distance duality relation $\eta(z)$ and evaluate its properties with minimal assumptions about the cosmological model, relying instead on the observational data and the GP framework.

\begin{figure}
	\centering
	\includegraphics[scale=0.47]{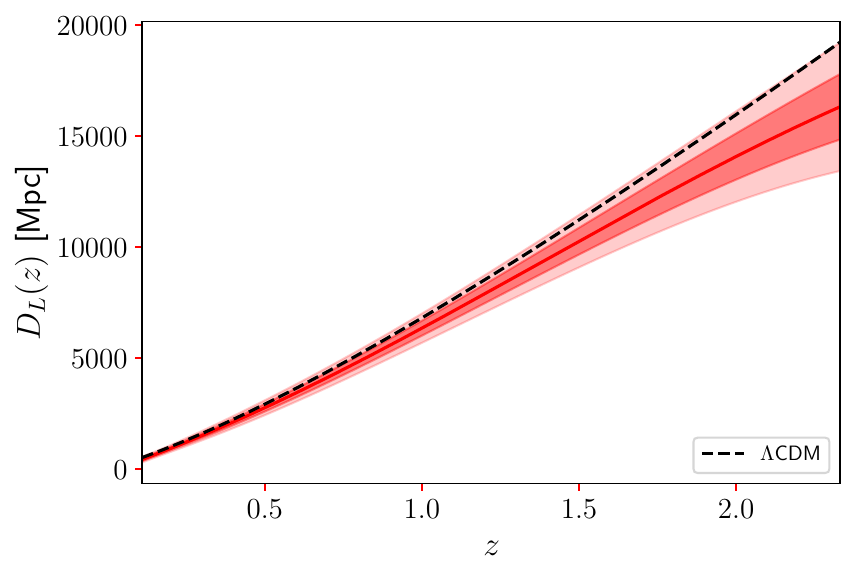}
	\includegraphics[scale=0.47]{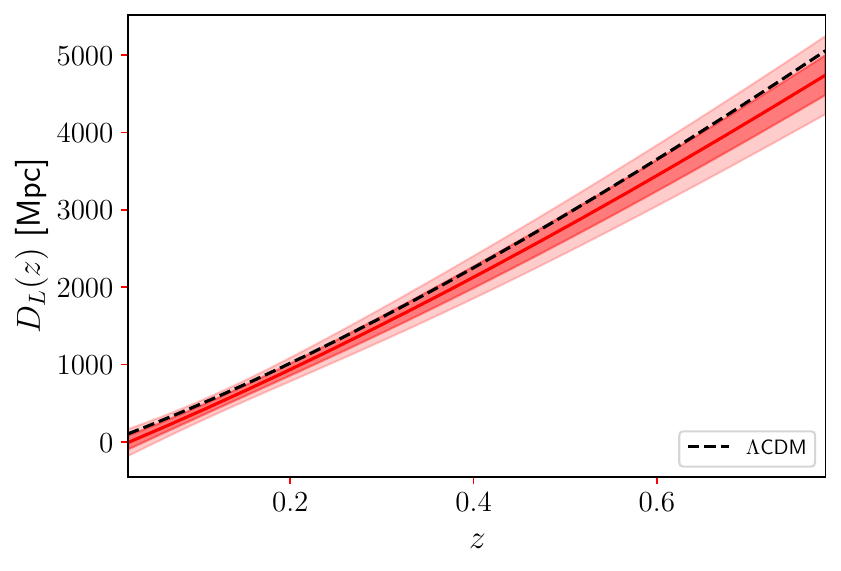}
    \caption{Reconstruction of the luminosity distance, $D_L(z)$, using our final dataset comprising 2861 measurements (1701 SNIa and 1160 QSO) binned in 25 data points in the interval $0<z<2.5$ with $\Delta z=0.1$. The solid red curve represents the mean Gaussian Process reconstruction, while the light and dark shaded regions correspond to the $1\sigma$ and $2\sigma$ confidence intervals, respectively. Upper panel: Reconstruction in the interval $0.11\leq z \leq 2.33$. Lower panel: Reconstruction in the interval $0.023\leq z \leq 0.784$.}
	\label{DL_rec}
\end{figure}

\section{Results}
\label{sec5:results}

\begin{figure}[htp]
	\centering
	\includegraphics[scale=0.47]{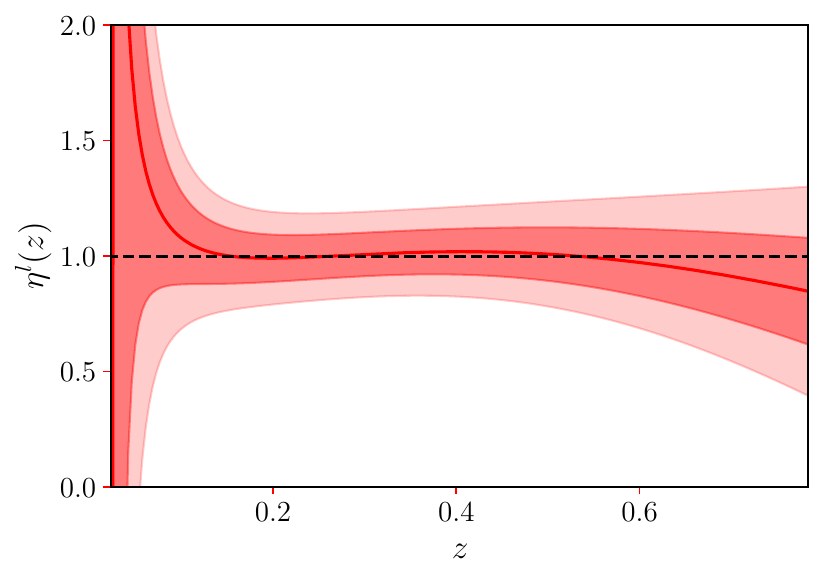} \\
	\includegraphics[scale=0.47]{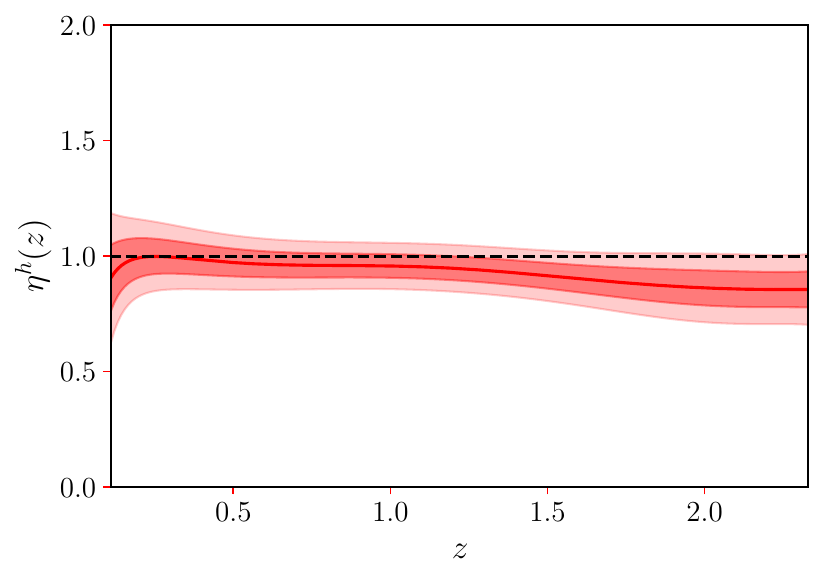} \\
	\includegraphics[scale=0.47]{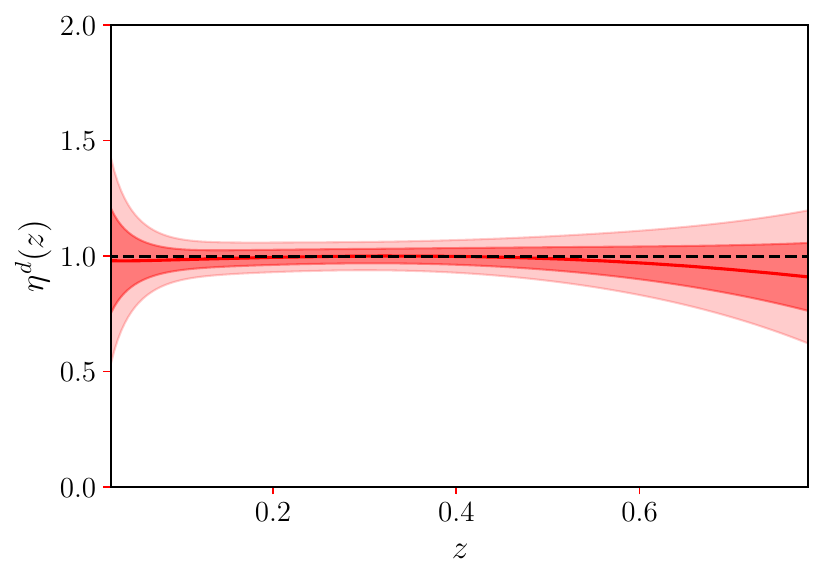}
\caption{These plots show the statistical GP reconstruction of the functions $\eta^{\,l, \,h,\,d}(z)$, 
with their corresponding $1\sigma$ and $2\sigma$ confidence intervals. 
For comparison, the dashed line corresponds to the constant value $\eta = 1$.
    The three panels correspond to different data combinations: The top panel represents the low-redshift combination ($\eta^{\,l}$), obtained by combining the GC measurements of $D_A(z)$ with the full SNIa + QSO sample for $D_L(z)$. The middle panel shows the high-redshift combination ($\eta^{\,h}$), based on the BAO-derived $D_A(z)$ values combined with the SNIa + QSO data. Finally, the bottom panel illustrates the distance dual angular probe combination ($\eta^{\,d}$), resulting from the combination of GC and BAO data alone, without involving luminosity distances.}
	\label{fig:eta_recon}
\end{figure}

In Figure \ref{fig:DA_recon}, we show the GP reconstruction of $D_A(z)$ for GC and BAO, in the upper and lower panels, respectively.  We reconstruct the GC data in the redshift interval $0.023 \leq z \leq 0.784$ and the BAO data in the redshift interval $0.11 \leq z \leq 2.33$.  Later, we adapt the BAO reconstruction interval to the same 
redshift range of the GC data, in order to reconstruct $\eta(z)$ using equation (\ref{eq:eta_da_modified}). It is important to note that the inferences for $D_{A,\rm BAO}$ are significantly more precise than those for $D_{A,\rm GC}$ when evaluated over the same redshift range, for example, $0.02 \leq z \leq 0.8$. This difference will become evident in the discussion that follows.
Therefore, the inclusion of the GC sample in this work serves primarily in the context of performing a statistical reconstruction of $\eta(z)$ using different datasets (including systematics) and for a simple qualitative comparison. In terms of precision, however, the BAO sample is considerably more accurate than the GC sample and provides robust statistical information on its own. However, note that we assume the fiducial value for $r_d$ that is calculated assuming the flat $\Lambda$CDM model. Given the uncertainty in our measurements, a 5\% variation in the value of $r_d$ has no significant impact on our result.

Regarding the $D_L$ data, we have compiled a total of 3896 
measurements, comprising 2195 QSO and 1701 SNIa, covering the redshift interval $0.001 \leq z \leq 2.5$. 
These observational data were then compared with the theoretical predictions of the $\Lambda$CDM model. 
In the upper panel of Figure~\ref{fig:mu_data_cut_DL_bin}, 
the gray squares represent the residuals of the QSO data with respect to the $\Lambda$CDM model, while the blue dots correspond to the SNIa residuals. 
The red dashed lines mark the $1 \sigma$ dispersion of the SNIa distribution, providing a reference for the typical uncertainty range. 

To improve the quality of the reconstruction and reduce the impact of outliers, we excluded QSO data falling outside the $1\sigma$ SNIa dispersion band. 
This filtering resulted in a refined sample containing 
1701 SNIa and 1160 QSO, a total of 2861 $D_L$ measurements 
which are used for the subsequent statistical analysis. 
In fact, the lower panel of Figure~\ref{fig:mu_data_cut_DL_bin} shows the binned data points (red squares) for this sample of 2861 data. 
These binned values serve as inputs for the Gaussian Process regression in the next stage of the analysis. The binning interval was $0<z<2.5$ with $\Delta z=0.1$. The interval was chosen after successive tests. It cannot be too large, otherwise the distribution would be affected by distance evolution, nor too small, as the noise would become dominant. Note that the methodology depends on the cosmological parameters of the Flat-$\Lambda$CDM model, in particular on the density parameters $\Omega_{m,0}$ e $\Omega_{k}$.
Figure~\ref{DL_rec} shows the Gaussian Process statistical reconstruction of the luminosity distance, $D_L(z)$, based on our final selected sample consisting of 2861 data. 
Our reconstruction provides a non-parametric estimate, and the corresponding confidence intervals, of $D_L(z)$ over the redshift interval $0.11 \leq z \leq 2.33$ to match the BAO $z$-interval; 
and, on the other hand, the second reconstruction of $D_L(z)$ is done in the redshift interval $0.023 \leq z \leq 0.784$ to match the GC $z$-interval. 

With the $D_A(z)$ and $D_L(z)$ reconstructed functions
in hand, we shall test the validity of the CDDR. 
In this study, we quantify potential deviations from the standard CDDR by introducing the parameter $\eta(z)$, 
defined as
\begin{equation}
	\label{eq:eta_z}
	\eta(z) \equiv \frac{D_L(z)}{D_A(z)}(1+z)^{-2} \,,
\end{equation}
where $\eta(z) = 1$ corresponds to the perfect validity of the CDDR. 

Figure~\ref{fig:eta_recon} displays the reconstructed evolution of the CDDR parameter $\eta(z)$, derived according to Equation~(\ref{eq:eta_z}). 
In this figure, we present three GP reconstructions based on specific combinations of angular and luminosity distance data, as explained below: 

\begin{itemize}
	\item Low redshift combination ($\eta^{\,l}$): obtained by combining the galaxy cluster (GC) measurements of $D_A(z)$ with the full SNIa + QSO sample for $D_L(z)$; 
	both datasets for the redshift interval $0.023 \le z \le 0.784$.
	\item High redshift combination ($\eta^{\,h}$): based on the BAO-derived $D_A(z)$ values combined with the SNIa + QSO data; both datasets for the redshift interval 
	$0.023 \le z \le 2.33$.
	\item Distance dual angular probe combination ($\eta^{\,d}$): resulting from the combination of GC and BAO data alone, without involving luminosity distances; 
	both datasets for the redshift interval $0.023 \le z \le 0.784$.
\end{itemize}

In all three cases, the solid curves represent the mean reconstructions of $\eta(z)$, while the shaded bands denote the $1 \sigma$ and $2 \sigma$ confidence regions. 
Our analyses indicate no statistically significant deviations from the standard CDDR, i.e., $\eta(z) = 1$, within the redshift ranges investigated. 
All reconstructed values are consistent with unity at the 
$2 \sigma$ level. 

These results reinforce the robustness of the CDDR under current observational constraints, including high redshift cosmological probes. 
Nevertheless, future improvements in the precision and systematics of $D_L$ measurements from high-redshift QSO will play a crucial role in enabling more precise and extended tests of the CDDR across a wider redshift domain.

    \section{Conclusions}
\label{sec6:conclusions}

In this work, we probed the validity of the cosmic distance duality relation, CDDR, using non-parametric GP reconstruction for high redshift cosmological data. 

This fundamental relation holds in any metric theory of gravity under the assumptions of photon number conservation and that photons travel along null geodesics. To perform this test, we employed a combination of $D_A$ measurements from transverse BAO phenomena and GC features, along with $D_L$ estimates from SNIa and QSO, as presented in 
section~\ref{sec2:data}. 
Given the large intrinsic scatter in QSO luminosity distances, we implemented a data selection procedure based on their deviations from the predictions of the fiducial 
$\Lambda$CDM model. 
We retained QSO consistent with SNIa data within a $1\sigma$ 
uncertainty region, then a binning technique was applied to further reduce statistical noise. 
This resulted in a refined high-quality sample, consisting of 1701 SNIa and 1160 QSO measurements of $D_L$ 
spanning the redshift range $0.001 \leq z \leq 2.5$. We emphasize that the quasar filtering process explicitly assumes the $\Lambda$CDM model. It is expected that in the future the dispersion of quasars will decrease and thus it will not be necessary to use model-dependent mechanisms to reduce noise in cosmological analyses.

Our main results covers the redshift interval $0.023 \leq z \leq 2.33$, where we combine three complementary pairings of angular and luminosity distance datasets: GC with SNIa + QSO at low redshifts, BAO with SNIa + QSO at intermediate-to-high redshifts, and GC with BAO for consistency cross-checks. 
The reconstructed $\eta(z)$ shows no statistically significant deviation from the null hypothesis. 
These findings support the validity of the CDDR under current observational constraints. 
In addition, our study demonstrate the reliability of using QSO data, once properly filtered and binned, for precision cosmology.

Our methodology underscores the power of combining multiple cosmological probes across a broad redshift range to test fundamental assumptions of the standard model. 
In particular, the inclusion of high-redshift QSO allows us to extend the reach of CDDR tests well beyond the limits of SNIa-based analyses. 
Future advances in QSO and SNIa luminosity distance measurements, both in precision and redshift coverage, will allow even more stringent and model-independent tests of the CDDR. Enabling access to the robustness of our analyses regarding dependence on priors, as in the case of $r_d$, and in the process of filtering quasars, which depends on the flat $\Lambda$CDM.
This may open the possibility of detecting new physics, such as photon-axion mixing, violations of photon conservation, or non-metric gravitational effects, if statistically significant deviations are observed.


\begin{acknowledgments}
FA thanks to Funda\c{c}\~{a}o Carlos Chagas Filho de Amparo \`{a} Pesquisa do Estado do Rio de Janeiro (FAPERJ), Processo SEI-260003/001221/2025, for the financial support.
FO, CF, and ML thank CAPES for the fellowship. R.C.N. thanks the financial support from the Conselho Nacional de Desenvolvimento Científico e Tecnologico (CNPq, National Council for Scientific and Technological Development) under the project No. 304306/2022-3, and the Fundação de Amparo à Pesquisa do Estado do RS (FAPERGS, Research Support Foundation of the State of RS) for partial financial support under the project No. 23/2551-0000848-3. AB acknowledges a CNPq fellowship.  RFLH thanks to CNPq support under the project No.308550/2023-4.
\end{acknowledgments}

\section*{Data availability}

Data will be made available on request.

\bibliography{main}

\begin{thebibliography}{118}%
\makeatletter
\providecommand \@ifxundefined [1]{%
 \@ifx{#1\undefined}
}%
\providecommand \@ifnum [1]{%
 \ifnum #1\expandafter \@firstoftwo
 \else \expandafter \@secondoftwo
 \fi
}%
\providecommand \@ifx [1]{%
 \ifx #1\expandafter \@firstoftwo
 \else \expandafter \@secondoftwo
 \fi
}%
\providecommand \natexlab [1]{#1}%
\providecommand \enquote  [1]{``#1''}%
\providecommand \bibnamefont  [1]{#1}%
\providecommand \bibfnamefont [1]{#1}%
\providecommand \citenamefont [1]{#1}%
\providecommand \href@noop [0]{\@secondoftwo}%
\providecommand \href [0]{\begingroup \@sanitize@url \@href}%
\providecommand \@href[1]{\@@startlink{#1}\@@href}%
\providecommand \@@href[1]{\endgroup#1\@@endlink}%
\providecommand \@sanitize@url [0]{\catcode `\\12\catcode `\$12\catcode
  `\&12\catcode `\#12\catcode `\^12\catcode `\_12\catcode `\%12\relax}%
\providecommand \@@startlink[1]{}%
\providecommand \@@endlink[0]{}%
\providecommand \url  [0]{\begingroup\@sanitize@url \@url }%
\providecommand \@url [1]{\endgroup\@href {#1}{\urlprefix }}%
\providecommand \urlprefix  [0]{URL }%
\providecommand \Eprint [0]{\href }%
\providecommand \doibase [0]{http://dx.doi.org/}%
\providecommand \selectlanguage [0]{\@gobble}%
\providecommand \bibinfo  [0]{\@secondoftwo}%
\providecommand \bibfield  [0]{\@secondoftwo}%
\providecommand \translation [1]{[#1]}%
\providecommand \BibitemOpen [0]{}%
\providecommand \bibitemStop [0]{}%
\providecommand \bibitemNoStop [0]{.\EOS\space}%
\providecommand \EOS [0]{\spacefactor3000\relax}%
\providecommand \BibitemShut  [1]{\csname bibitem#1\endcsname}%
\let\auto@bib@innerbib\@empty
\bibitem [{\citenamefont {Di~Valentino}\ \emph {et~al.}(2025)\citenamefont
  {Di~Valentino} \emph {et~al.}}]{DiValentino:2025sru}%
  \BibitemOpen
  \bibfield  {author} {\bibinfo {author} {\bibfnamefont {E.}~\bibnamefont
  {Di~Valentino}} \emph {et~al.},\ }\href@noop {} {\  (\bibinfo {year}
  {2025})},\ \Eprint {http://arxiv.org/abs/2504.01669} {arXiv:2504.01669
  [astro-ph.CO]} \BibitemShut {NoStop}%
\bibitem [{\citenamefont {{Caldwell}}\ and\ \citenamefont
  {{Kamionkowski}}(2009)}]{2009ARNPS..59..397C}%
  \BibitemOpen
  \bibfield  {author} {\bibinfo {author} {\bibfnamefont {R.~R.}\ \bibnamefont
  {{Caldwell}}}\ and\ \bibinfo {author} {\bibfnamefont {M.}~\bibnamefont
  {{Kamionkowski}}},\ }\href {\doibase 10.1146/annurev-nucl-010709-151330}
  {\bibfield  {journal} {\bibinfo  {journal} {Annual Review of Nuclear and
  Particle Science}\ }\textbf {\bibinfo {volume} {59}},\ \bibinfo {pages} {397}
  (\bibinfo {year} {2009})},\ \Eprint {http://arxiv.org/abs/0903.0866}
  {arXiv:0903.0866 [astro-ph.CO]} \BibitemShut {NoStop}%
\bibitem [{\citenamefont {{Weinberg}}\ \emph {et~al.}(2013)\citenamefont
  {{Weinberg}}, \citenamefont {{Mortonson}}, \citenamefont {{Eisenstein}},
  \citenamefont {{Hirata}}, \citenamefont {{Riess}},\ and\ \citenamefont
  {{Rozo}}}]{2013PhR...530...87W}%
  \BibitemOpen
  \bibfield  {author} {\bibinfo {author} {\bibfnamefont {D.~H.}\ \bibnamefont
  {{Weinberg}}}, \bibinfo {author} {\bibfnamefont {M.~J.}\ \bibnamefont
  {{Mortonson}}}, \bibinfo {author} {\bibfnamefont {D.~J.}\ \bibnamefont
  {{Eisenstein}}}, \bibinfo {author} {\bibfnamefont {C.}~\bibnamefont
  {{Hirata}}}, \bibinfo {author} {\bibfnamefont {A.~G.}\ \bibnamefont
  {{Riess}}}, \ and\ \bibinfo {author} {\bibfnamefont {E.}~\bibnamefont
  {{Rozo}}},\ }\href {\doibase 10.1016/j.physrep.2013.05.001} {\bibfield
  {journal} {\bibinfo  {journal} {Phys. Rep.}\ }\textbf {\bibinfo {volume}
  {530}},\ \bibinfo {pages} {87} (\bibinfo {year} {2013})},\ \Eprint
  {http://arxiv.org/abs/1201.2434} {arXiv:1201.2434 [astro-ph.CO]} \BibitemShut
  {NoStop}%
\bibitem [{\citenamefont {{Perivolaropoulos}}\ and\ \citenamefont
  {{Skara}}(2022)}]{2022NewAR..9501659P}%
  \BibitemOpen
  \bibfield  {author} {\bibinfo {author} {\bibfnamefont {L.}~\bibnamefont
  {{Perivolaropoulos}}}\ and\ \bibinfo {author} {\bibfnamefont
  {F.}~\bibnamefont {{Skara}}},\ }\href {\doibase 10.1016/j.newar.2022.101659}
  {\bibfield  {journal} {\bibinfo  {journal} {nar}\ }\textbf {\bibinfo {volume}
  {95}},\ \bibinfo {eid} {101659} (\bibinfo {year} {2022})},\ \Eprint
  {http://arxiv.org/abs/2105.05208} {arXiv:2105.05208 [astro-ph.CO]}
  \BibitemShut {NoStop}%
\bibitem [{\citenamefont {Vagnozzi}(2023)}]{Vagnozzi:2023nrq}%
  \BibitemOpen
  \bibfield  {author} {\bibinfo {author} {\bibfnamefont {S.}~\bibnamefont
  {Vagnozzi}},\ }\href {\doibase 10.3390/universe9090393} {\bibfield  {journal}
  {\bibinfo  {journal} {Universe}\ }\textbf {\bibinfo {volume} {9}},\ \bibinfo
  {pages} {393} (\bibinfo {year} {2023})},\ \Eprint
  {http://arxiv.org/abs/2308.16628} {arXiv:2308.16628 [astro-ph.CO]}
  \BibitemShut {NoStop}%
\bibitem [{\citenamefont {{Etherington}}(1933)}]{Etherington33}%
  \BibitemOpen
  \bibfield  {author} {\bibinfo {author} {\bibfnamefont {I.~M.~H.}\
  \bibnamefont {{Etherington}}},\ }\href@noop {} {\bibfield  {journal}
  {\bibinfo  {journal} {Philosophical Magazine}\ }\textbf {\bibinfo {volume}
  {15}},\ \bibinfo {pages} {761} (\bibinfo {year} {1933})}\BibitemShut
  {NoStop}%
\bibitem [{\citenamefont {{Ellis}}(2007)}]{Ellis07}%
  \BibitemOpen
  \bibfield  {author} {\bibinfo {author} {\bibfnamefont {G.~F.~R.}\
  \bibnamefont {{Ellis}}},\ }\href {\doibase 10.1007/s10714-006-0355-5}
  {\bibfield  {journal} {\bibinfo  {journal} {General Relativity and
  Gravitation}\ }\textbf {\bibinfo {volume} {39}},\ \bibinfo {pages} {1047}
  (\bibinfo {year} {2007})}\BibitemShut {NoStop}%
\bibitem [{\citenamefont {{Qin}}\ \emph {et~al.}(2021)\citenamefont {{Qin}},
  \citenamefont {{Melia}},\ and\ \citenamefont {{Zhang}}}]{Qin21}%
  \BibitemOpen
  \bibfield  {author} {\bibinfo {author} {\bibfnamefont {J.}~\bibnamefont
  {{Qin}}}, \bibinfo {author} {\bibfnamefont {F.}~\bibnamefont {{Melia}}}, \
  and\ \bibinfo {author} {\bibfnamefont {T.-J.}\ \bibnamefont {{Zhang}}},\
  }\href {\doibase 10.1093/mnras/stab124} {\bibfield  {journal} {\bibinfo
  {journal} {MNRAS}\ }\textbf {\bibinfo {volume} {502}},\ \bibinfo {pages}
  {3500} (\bibinfo {year} {2021})},\ \Eprint {http://arxiv.org/abs/2101.05574}
  {arXiv:2101.05574 [astro-ph.CO]} \BibitemShut {NoStop}%
\bibitem [{\citenamefont {{Bassett}}\ and\ \citenamefont
  {{Kunz}}(2004)}]{Bassett04}%
  \BibitemOpen
  \bibfield  {author} {\bibinfo {author} {\bibfnamefont {B.~A.}\ \bibnamefont
  {{Bassett}}}\ and\ \bibinfo {author} {\bibfnamefont {M.}~\bibnamefont
  {{Kunz}}},\ }\href {\doibase 10.1103/PhysRevD.69.101305} {\bibfield
  {journal} {\bibinfo  {journal} {PRD}\ }\textbf {\bibinfo {volume} {69}},\
  \bibinfo {eid} {101305} (\bibinfo {year} {2004})},\ \Eprint
  {http://arxiv.org/abs/astro-ph/0312443} {arXiv:astro-ph/0312443 [astro-ph]}
  \BibitemShut {NoStop}%
\bibitem [{\citenamefont {{Hees}}\ \emph {et~al.}(2014)\citenamefont {{Hees}},
  \citenamefont {{Minazzoli}},\ and\ \citenamefont
  {{Larena}}}]{2014PhRvD..90l4064H}%
  \BibitemOpen
  \bibfield  {author} {\bibinfo {author} {\bibfnamefont {A.}~\bibnamefont
  {{Hees}}}, \bibinfo {author} {\bibfnamefont {O.}~\bibnamefont {{Minazzoli}}},
  \ and\ \bibinfo {author} {\bibfnamefont {J.}~\bibnamefont {{Larena}}},\
  }\href {\doibase 10.1103/PhysRevD.90.124064} {\bibfield  {journal} {\bibinfo
  {journal} {PRD}\ }\textbf {\bibinfo {volume} {90}},\ \bibinfo {eid} {124064}
  (\bibinfo {year} {2014})},\ \Eprint {http://arxiv.org/abs/1406.6187}
  {arXiv:1406.6187 [astro-ph.CO]} \BibitemShut {NoStop}%
\bibitem [{\citenamefont {{Azevedo}}\ and\ \citenamefont
  {{Avelino}}(2021)}]{2021PhRvD.104h4079A}%
  \BibitemOpen
  \bibfield  {author} {\bibinfo {author} {\bibfnamefont {R.~P.~L.}\
  \bibnamefont {{Azevedo}}}\ and\ \bibinfo {author} {\bibfnamefont {P.~P.}\
  \bibnamefont {{Avelino}}},\ }\href {\doibase 10.1103/PhysRevD.104.084079}
  {\bibfield  {journal} {\bibinfo  {journal} {PRD}\ }\textbf {\bibinfo {volume}
  {104}},\ \bibinfo {eid} {084079} (\bibinfo {year} {2021})},\ \Eprint
  {http://arxiv.org/abs/2104.01209} {arXiv:2104.01209 [gr-qc]} \BibitemShut
  {NoStop}%
\bibitem [{\citenamefont {{Levi Said}}\ \emph {et~al.}(2020)\citenamefont
  {{Levi Said}}, \citenamefont {{Mifsud}}, \citenamefont {{Parkinson}},
  \citenamefont {{Saridakis}}, \citenamefont {{Sultana}},\ and\ \citenamefont
  {{Zarb Adami}}}]{2020JCAP...11..047L}%
  \BibitemOpen
  \bibfield  {author} {\bibinfo {author} {\bibfnamefont {J.}~\bibnamefont
  {{Levi Said}}}, \bibinfo {author} {\bibfnamefont {J.}~\bibnamefont
  {{Mifsud}}}, \bibinfo {author} {\bibfnamefont {D.}~\bibnamefont
  {{Parkinson}}}, \bibinfo {author} {\bibfnamefont {E.~N.}\ \bibnamefont
  {{Saridakis}}}, \bibinfo {author} {\bibfnamefont {J.}~\bibnamefont
  {{Sultana}}}, \ and\ \bibinfo {author} {\bibfnamefont {K.}~\bibnamefont
  {{Zarb Adami}}},\ }\href {\doibase 10.1088/1475-7516/2020/11/047} {\bibfield
  {journal} {\bibinfo  {journal} {JCAP}\ }\textbf {\bibinfo {volume} {2020}},\
  \bibinfo {eid} {047} (\bibinfo {year} {2020})},\ \Eprint
  {http://arxiv.org/abs/2005.05368} {arXiv:2005.05368 [astro-ph.CO]}
  \BibitemShut {NoStop}%
\bibitem [{\citenamefont {{Holanda}}\ and\ \citenamefont
  {{Barros}}(2016)}]{2016PhRvD..94b3524H}%
  \BibitemOpen
  \bibfield  {author} {\bibinfo {author} {\bibfnamefont {R.~F.~L.}\
  \bibnamefont {{Holanda}}}\ and\ \bibinfo {author} {\bibfnamefont
  {K.~N.~N.~O.}\ \bibnamefont {{Barros}}},\ }\href {\doibase
  10.1103/PhysRevD.94.023524} {\bibfield  {journal} {\bibinfo  {journal} {PRD}\
  }\textbf {\bibinfo {volume} {94}},\ \bibinfo {eid} {023524} (\bibinfo {year}
  {2016})},\ \Eprint {http://arxiv.org/abs/1606.07923} {arXiv:1606.07923
  [astro-ph.CO]} \BibitemShut {NoStop}%
\bibitem [{\citenamefont {{Holanda}}\ \emph {et~al.}(2017)\citenamefont
  {{Holanda}}, \citenamefont {{Pereira}}, \citenamefont {{Busti}},\ and\
  \citenamefont {{Bessa}}}]{2017CQGra..34s5003H}%
  \BibitemOpen
  \bibfield  {author} {\bibinfo {author} {\bibfnamefont {R.~F.~L.}\
  \bibnamefont {{Holanda}}}, \bibinfo {author} {\bibfnamefont {S.~H.}\
  \bibnamefont {{Pereira}}}, \bibinfo {author} {\bibfnamefont {V.~C.}\
  \bibnamefont {{Busti}}}, \ and\ \bibinfo {author} {\bibfnamefont {C.~H.~G.}\
  \bibnamefont {{Bessa}}},\ }\href {\doibase 10.1088/1361-6382/aa8828}
  {\bibfield  {journal} {\bibinfo  {journal} {Classical and Quantum Gravity}\
  }\textbf {\bibinfo {volume} {34}},\ \bibinfo {eid} {195003} (\bibinfo {year}
  {2017})},\ \Eprint {http://arxiv.org/abs/1705.05439} {arXiv:1705.05439
  [gr-qc]} \BibitemShut {NoStop}%
\bibitem [{\citenamefont {{Kumar}}\ \emph {et~al.}(2022)\citenamefont
  {{Kumar}}, \citenamefont {{Rana}}, \citenamefont {{Jain}}, \citenamefont
  {{Mahajan}}, \citenamefont {{Mukherjee}},\ and\ \citenamefont
  {{Holanda}}}]{2022JCAP...01..053K}%
  \BibitemOpen
  \bibfield  {author} {\bibinfo {author} {\bibfnamefont {D.}~\bibnamefont
  {{Kumar}}}, \bibinfo {author} {\bibfnamefont {A.}~\bibnamefont {{Rana}}},
  \bibinfo {author} {\bibfnamefont {D.}~\bibnamefont {{Jain}}}, \bibinfo
  {author} {\bibfnamefont {S.}~\bibnamefont {{Mahajan}}}, \bibinfo {author}
  {\bibfnamefont {A.}~\bibnamefont {{Mukherjee}}}, \ and\ \bibinfo {author}
  {\bibfnamefont {R.~F.~L.}\ \bibnamefont {{Holanda}}},\ }\href {\doibase
  10.1088/1475-7516/2022/01/053} {\bibfield  {journal} {\bibinfo  {journal}
  {JCAP}\ }\textbf {\bibinfo {volume} {2022}},\ \bibinfo {eid} {053} (\bibinfo
  {year} {2022})},\ \Eprint {http://arxiv.org/abs/2107.04784} {arXiv:2107.04784
  [astro-ph.CO]} \BibitemShut {NoStop}%
\bibitem [{\citenamefont {{Holanda}}\ \emph {et~al.}(2018)\citenamefont
  {{Holanda}}, \citenamefont {{Pereira}},\ and\ \citenamefont
  {{Jain}}}]{2018PhRvD..97b3538H}%
  \BibitemOpen
  \bibfield  {author} {\bibinfo {author} {\bibfnamefont {R.~F.~L.}\
  \bibnamefont {{Holanda}}}, \bibinfo {author} {\bibfnamefont {S.~H.}\
  \bibnamefont {{Pereira}}}, \ and\ \bibinfo {author} {\bibfnamefont
  {D.}~\bibnamefont {{Jain}}},\ }\href {\doibase 10.1103/PhysRevD.97.023538}
  {\bibfield  {journal} {\bibinfo  {journal} {PRD}\ }\textbf {\bibinfo {volume}
  {97}},\ \bibinfo {eid} {023538} (\bibinfo {year} {2018})},\ \Eprint
  {http://arxiv.org/abs/1801.04344} {arXiv:1801.04344 [astro-ph.CO]}
  \BibitemShut {NoStop}%
\bibitem [{\citenamefont {{Combes}}(2004)}]{2004NewAR..48..583C}%
  \BibitemOpen
  \bibfield  {author} {\bibinfo {author} {\bibfnamefont {F.}~\bibnamefont
  {{Combes}}},\ }\href {\doibase 10.1016/j.newar.2003.12.053} {\bibfield
  {journal} {\bibinfo  {journal} {nar}\ }\textbf {\bibinfo {volume} {48}},\
  \bibinfo {pages} {583} (\bibinfo {year} {2004})},\ \Eprint
  {http://arxiv.org/abs/astro-ph/0308144} {arXiv:astro-ph/0308144 [astro-ph]}
  \BibitemShut {NoStop}%
\bibitem [{\citenamefont {{Shafieloo}}\ \emph {et~al.}(2013)\citenamefont
  {{Shafieloo}}, \citenamefont {{Majumdar}}, \citenamefont {{Sahni}},\ and\
  \citenamefont {{Starobinsky}}}]{2013JCAP...04..042S}%
  \BibitemOpen
  \bibfield  {author} {\bibinfo {author} {\bibfnamefont {A.}~\bibnamefont
  {{Shafieloo}}}, \bibinfo {author} {\bibfnamefont {S.}~\bibnamefont
  {{Majumdar}}}, \bibinfo {author} {\bibfnamefont {V.}~\bibnamefont {{Sahni}}},
  \ and\ \bibinfo {author} {\bibfnamefont {A.~A.}\ \bibnamefont
  {{Starobinsky}}},\ }\href {\doibase 10.1088/1475-7516/2013/04/042} {\bibfield
   {journal} {\bibinfo  {journal} {JCAP}\ }\textbf {\bibinfo {volume} {2013}},\
  \bibinfo {eid} {042} (\bibinfo {year} {2013})},\ \Eprint
  {http://arxiv.org/abs/1212.1277} {arXiv:1212.1277 [astro-ph.CO]} \BibitemShut
  {NoStop}%
\bibitem [{\citenamefont {{Jesus}}\ \emph {et~al.}(2017)\citenamefont
  {{Jesus}}, \citenamefont {{Holanda}},\ and\ \citenamefont
  {{Dantas}}}]{2017GReGr..49..150J}%
  \BibitemOpen
  \bibfield  {author} {\bibinfo {author} {\bibfnamefont {J.~F.}\ \bibnamefont
  {{Jesus}}}, \bibinfo {author} {\bibfnamefont {R.~F.~L.}\ \bibnamefont
  {{Holanda}}}, \ and\ \bibinfo {author} {\bibfnamefont {M.~A.}\ \bibnamefont
  {{Dantas}}},\ }\href {\doibase 10.1007/s10714-017-2317-5} {\bibfield
  {journal} {\bibinfo  {journal} {General Relativity and Gravitation}\ }\textbf
  {\bibinfo {volume} {49}},\ \bibinfo {eid} {150} (\bibinfo {year} {2017})},\
  \Eprint {http://arxiv.org/abs/1605.01342} {arXiv:1605.01342 [astro-ph.CO]}
  \BibitemShut {NoStop}%
\bibitem [{\citenamefont {{Li}}\ \emph {et~al.}(2013)\citenamefont {{Li}},
  \citenamefont {{Wu}}, \citenamefont {{Yu}},\ and\ \citenamefont
  {{Zhu}}}]{2013PhRvD..87j3013L}%
  \BibitemOpen
  \bibfield  {author} {\bibinfo {author} {\bibfnamefont {Z.}~\bibnamefont
  {{Li}}}, \bibinfo {author} {\bibfnamefont {P.}~\bibnamefont {{Wu}}}, \bibinfo
  {author} {\bibfnamefont {H.}~\bibnamefont {{Yu}}}, \ and\ \bibinfo {author}
  {\bibfnamefont {Z.-H.}\ \bibnamefont {{Zhu}}},\ }\href {\doibase
  10.1103/PhysRevD.87.103013} {\bibfield  {journal} {\bibinfo  {journal} {PRD}\
  }\textbf {\bibinfo {volume} {87}},\ \bibinfo {eid} {103013} (\bibinfo {year}
  {2013})},\ \Eprint {http://arxiv.org/abs/1304.7317} {arXiv:1304.7317
  [astro-ph.CO]} \BibitemShut {NoStop}%
\bibitem [{\citenamefont {{Liao}}\ \emph {et~al.}(2013)\citenamefont {{Liao}},
  \citenamefont {{Li}}, \citenamefont {{Ming}},\ and\ \citenamefont
  {{Zhu}}}]{2013PhLB..718.1166L}%
  \BibitemOpen
  \bibfield  {author} {\bibinfo {author} {\bibfnamefont {K.}~\bibnamefont
  {{Liao}}}, \bibinfo {author} {\bibfnamefont {Z.}~\bibnamefont {{Li}}},
  \bibinfo {author} {\bibfnamefont {J.}~\bibnamefont {{Ming}}}, \ and\ \bibinfo
  {author} {\bibfnamefont {Z.-H.}\ \bibnamefont {{Zhu}}},\ }\href {\doibase
  10.1016/j.physletb.2012.12.022} {\bibfield  {journal} {\bibinfo  {journal}
  {Physics Letters B}\ }\textbf {\bibinfo {volume} {718}},\ \bibinfo {pages}
  {1166} (\bibinfo {year} {2013})},\ \Eprint {http://arxiv.org/abs/1212.6612}
  {arXiv:1212.6612 [astro-ph.CO]} \BibitemShut {NoStop}%
\bibitem [{\citenamefont {Rasanen}\ \emph {et~al.}(2016)\citenamefont
  {Rasanen}, \citenamefont {Valiviita},\ and\ \citenamefont
  {Kosonen}}]{Rasanen:2015kca}%
  \BibitemOpen
  \bibfield  {author} {\bibinfo {author} {\bibfnamefont {S.}~\bibnamefont
  {Rasanen}}, \bibinfo {author} {\bibfnamefont {J.}~\bibnamefont {Valiviita}},
  \ and\ \bibinfo {author} {\bibfnamefont {V.}~\bibnamefont {Kosonen}},\ }\href
  {\doibase 10.1088/1475-7516/2016/04/050} {\bibfield  {journal} {\bibinfo
  {journal} {JCAP}\ }\textbf {\bibinfo {volume} {04}},\ \bibinfo {pages} {050}
  (\bibinfo {year} {2016})},\ \Eprint {http://arxiv.org/abs/1512.05346}
  {arXiv:1512.05346 [astro-ph.CO]} \BibitemShut {NoStop}%
\bibitem [{\citenamefont {Renzi}\ \emph {et~al.}(2022)\citenamefont {Renzi},
  \citenamefont {Hogg},\ and\ \citenamefont {Giar\`e}}]{Renzi:2021xii}%
  \BibitemOpen
  \bibfield  {author} {\bibinfo {author} {\bibfnamefont {F.}~\bibnamefont
  {Renzi}}, \bibinfo {author} {\bibfnamefont {N.~B.}\ \bibnamefont {Hogg}}, \
  and\ \bibinfo {author} {\bibfnamefont {W.}~\bibnamefont {Giar\`e}},\ }\href
  {\doibase 10.1093/mnras/stac1030} {\bibfield  {journal} {\bibinfo  {journal}
  {Mon. Not. Roy. Astron. Soc.}\ }\textbf {\bibinfo {volume} {513}},\ \bibinfo
  {pages} {4004} (\bibinfo {year} {2022})},\ \Eprint
  {http://arxiv.org/abs/2112.05701} {arXiv:2112.05701 [astro-ph.CO]}
  \BibitemShut {NoStop}%
\bibitem [{\citenamefont {Qi}\ \emph {et~al.}(2025)\citenamefont {Qi},
  \citenamefont {Jiang}, \citenamefont {Hou},\ and\ \citenamefont
  {Zhang}}]{Qi:2024acx}%
  \BibitemOpen
  \bibfield  {author} {\bibinfo {author} {\bibfnamefont {J.-Z.}\ \bibnamefont
  {Qi}}, \bibinfo {author} {\bibfnamefont {Y.-F.}\ \bibnamefont {Jiang}},
  \bibinfo {author} {\bibfnamefont {W.-T.}\ \bibnamefont {Hou}}, \ and\
  \bibinfo {author} {\bibfnamefont {X.}~\bibnamefont {Zhang}},\ }\href
  {\doibase 10.3847/1538-4357/ad9de4} {\bibfield  {journal} {\bibinfo
  {journal} {Astrophys. J.}\ }\textbf {\bibinfo {volume} {979}},\ \bibinfo
  {pages} {2} (\bibinfo {year} {2025})},\ \Eprint
  {http://arxiv.org/abs/2407.07336} {arXiv:2407.07336 [astro-ph.CO]}
  \BibitemShut {NoStop}%
\bibitem [{\citenamefont {Keil}\ \emph {et~al.}(2025)\citenamefont {Keil},
  \citenamefont {Nesseris}, \citenamefont {Tutusaus},\ and\ \citenamefont
  {Blanchard}}]{Keil:2025ysb}%
  \BibitemOpen
  \bibfield  {author} {\bibinfo {author} {\bibfnamefont {F.}~\bibnamefont
  {Keil}}, \bibinfo {author} {\bibfnamefont {S.}~\bibnamefont {Nesseris}},
  \bibinfo {author} {\bibfnamefont {I.}~\bibnamefont {Tutusaus}}, \ and\
  \bibinfo {author} {\bibfnamefont {A.}~\bibnamefont {Blanchard}},\ }\href@noop
  {} {\  (\bibinfo {year} {2025})},\ \Eprint {http://arxiv.org/abs/2504.01750}
  {arXiv:2504.01750 [astro-ph.CO]} \BibitemShut {NoStop}%
\bibitem [{\citenamefont {Rana}\ \emph {et~al.}(2016)\citenamefont {Rana},
  \citenamefont {Jain}, \citenamefont {Mahajan},\ and\ \citenamefont
  {Mukherjee}}]{Rana:2015feb}%
  \BibitemOpen
  \bibfield  {author} {\bibinfo {author} {\bibfnamefont {A.}~\bibnamefont
  {Rana}}, \bibinfo {author} {\bibfnamefont {D.}~\bibnamefont {Jain}}, \bibinfo
  {author} {\bibfnamefont {S.}~\bibnamefont {Mahajan}}, \ and\ \bibinfo
  {author} {\bibfnamefont {A.}~\bibnamefont {Mukherjee}},\ }\href {\doibase
  10.1088/1475-7516/2016/07/026} {\bibfield  {journal} {\bibinfo  {journal}
  {JCAP}\ }\textbf {\bibinfo {volume} {07}},\ \bibinfo {pages} {026} (\bibinfo
  {year} {2016})},\ \Eprint {http://arxiv.org/abs/1511.09223} {arXiv:1511.09223
  [astro-ph.CO]} \BibitemShut {NoStop}%
\bibitem [{\citenamefont {Gahlaut}(2025)}]{Gahlaut:2025lhv}%
  \BibitemOpen
  \bibfield  {author} {\bibinfo {author} {\bibfnamefont {S.}~\bibnamefont
  {Gahlaut}},\ }\href {\doibase 10.1088/1674-4527/adae45} {\bibfield  {journal}
  {\bibinfo  {journal} {Res. Astron. Astrophys.}\ }\textbf {\bibinfo {volume}
  {25}},\ \bibinfo {pages} {025019} (\bibinfo {year} {2025})},\ \Eprint
  {http://arxiv.org/abs/2501.15086} {arXiv:2501.15086 [gr-qc]} \BibitemShut
  {NoStop}%
\bibitem [{\citenamefont {Mukherjee}\ and\ \citenamefont
  {Mukherjee}(2021)}]{Mukherjee:2021kcu}%
  \BibitemOpen
  \bibfield  {author} {\bibinfo {author} {\bibfnamefont {P.}~\bibnamefont
  {Mukherjee}}\ and\ \bibinfo {author} {\bibfnamefont {A.}~\bibnamefont
  {Mukherjee}},\ }\href {\doibase 10.1093/mnras/stab1054} {\bibfield  {journal}
  {\bibinfo  {journal} {Mon. Not. Roy. Astron. Soc.}\ }\textbf {\bibinfo
  {volume} {504}},\ \bibinfo {pages} {3938} (\bibinfo {year} {2021})},\ \Eprint
  {http://arxiv.org/abs/2104.06066} {arXiv:2104.06066 [astro-ph.CO]}
  \BibitemShut {NoStop}%
\bibitem [{\citenamefont {More}\ \emph {et~al.}(2016)\citenamefont {More},
  \citenamefont {Niikura}, \citenamefont {Schneider}, \citenamefont
  {Schuller},\ and\ \citenamefont {Werner}}]{More:2016fca}%
  \BibitemOpen
  \bibfield  {author} {\bibinfo {author} {\bibfnamefont {S.}~\bibnamefont
  {More}}, \bibinfo {author} {\bibfnamefont {H.}~\bibnamefont {Niikura}},
  \bibinfo {author} {\bibfnamefont {J.}~\bibnamefont {Schneider}}, \bibinfo
  {author} {\bibfnamefont {F.~P.}\ \bibnamefont {Schuller}}, \ and\ \bibinfo
  {author} {\bibfnamefont {M.~C.}\ \bibnamefont {Werner}},\ }\href@noop {} {\
  (\bibinfo {year} {2016})},\ \Eprint {http://arxiv.org/abs/1612.08784}
  {arXiv:1612.08784 [astro-ph.CO]} \BibitemShut {NoStop}%
\bibitem [{\citenamefont {{Holanda}}\ \emph {et~al.}(2010)\citenamefont
  {{Holanda}}, \citenamefont {{Lima}},\ and\ \citenamefont
  {{Ribeiro}}}]{Holanda10}%
  \BibitemOpen
  \bibfield  {author} {\bibinfo {author} {\bibfnamefont {R.~F.~L.}\
  \bibnamefont {{Holanda}}}, \bibinfo {author} {\bibfnamefont {J.~A.~S.}\
  \bibnamefont {{Lima}}}, \ and\ \bibinfo {author} {\bibfnamefont {M.~B.}\
  \bibnamefont {{Ribeiro}}},\ }\href {\doibase 10.1088/2041-8205/722/2/L233}
  {\bibfield  {journal} {\bibinfo  {journal} {ApJL}\ }\textbf {\bibinfo
  {volume} {722}},\ \bibinfo {pages} {L233} (\bibinfo {year} {2010})},\ \Eprint
  {http://arxiv.org/abs/1005.4458} {arXiv:1005.4458 [astro-ph.CO]} \BibitemShut
  {NoStop}%
\bibitem [{\citenamefont {{Avgoustidis}}\ \emph {et~al.}(2010)\citenamefont
  {{Avgoustidis}}, \citenamefont {{Burrage}}, \citenamefont {{Redondo}},
  \citenamefont {{Verde}},\ and\ \citenamefont
  {{Jimenez}}}]{2010JCAP...10..024A}%
  \BibitemOpen
  \bibfield  {author} {\bibinfo {author} {\bibfnamefont {A.}~\bibnamefont
  {{Avgoustidis}}}, \bibinfo {author} {\bibfnamefont {C.}~\bibnamefont
  {{Burrage}}}, \bibinfo {author} {\bibfnamefont {J.}~\bibnamefont
  {{Redondo}}}, \bibinfo {author} {\bibfnamefont {L.}~\bibnamefont {{Verde}}},
  \ and\ \bibinfo {author} {\bibfnamefont {R.}~\bibnamefont {{Jimenez}}},\
  }\href {\doibase 10.1088/1475-7516/2010/10/024} {\bibfield  {journal}
  {\bibinfo  {journal} {JCAP}\ }\textbf {\bibinfo {volume} {2010}},\ \bibinfo
  {eid} {024} (\bibinfo {year} {2010})},\ \Eprint
  {http://arxiv.org/abs/1004.2053} {arXiv:1004.2053 [astro-ph.CO]} \BibitemShut
  {NoStop}%
\bibitem [{\citenamefont {{Holanda}}\ \emph {et~al.}(2016)\citenamefont
  {{Holanda}}, \citenamefont {{Busti}},\ and\ \citenamefont
  {{Alcaniz}}}]{2016JCAP...02..054H}%
  \BibitemOpen
  \bibfield  {author} {\bibinfo {author} {\bibfnamefont {R.~F.~L.}\
  \bibnamefont {{Holanda}}}, \bibinfo {author} {\bibfnamefont {V.~C.}\
  \bibnamefont {{Busti}}}, \ and\ \bibinfo {author} {\bibfnamefont {J.~S.}\
  \bibnamefont {{Alcaniz}}},\ }\href {\doibase 10.1088/1475-7516/2016/02/054}
  {\bibfield  {journal} {\bibinfo  {journal} {JCAP}\ }\textbf {\bibinfo
  {volume} {2016}},\ \bibinfo {pages} {054} (\bibinfo {year} {2016})},\ \Eprint
  {http://arxiv.org/abs/1512.02486} {arXiv:1512.02486 [astro-ph.CO]}
  \BibitemShut {NoStop}%
\bibitem [{\citenamefont {{Holanda}}\ \emph {et~al.}(2011)\citenamefont
  {{Holanda}}, \citenamefont {{Lima}},\ and\ \citenamefont
  {{Ribeiro}}}]{2011A&A...528L..14H}%
  \BibitemOpen
  \bibfield  {author} {\bibinfo {author} {\bibfnamefont {R.~F.~L.}\
  \bibnamefont {{Holanda}}}, \bibinfo {author} {\bibfnamefont {J.~A.~S.}\
  \bibnamefont {{Lima}}}, \ and\ \bibinfo {author} {\bibfnamefont {M.~B.}\
  \bibnamefont {{Ribeiro}}},\ }\href {\doibase 10.1051/0004-6361/201015547}
  {\bibfield  {journal} {\bibinfo  {journal} {aap}\ }\textbf {\bibinfo {volume}
  {528}},\ \bibinfo {eid} {L14} (\bibinfo {year} {2011})},\ \Eprint
  {http://arxiv.org/abs/1003.5906} {arXiv:1003.5906 [astro-ph.CO]} \BibitemShut
  {NoStop}%
\bibitem [{\citenamefont {{Holanda}}(2016)}]{2016APh....84...78H}%
  \BibitemOpen
  \bibfield  {author} {\bibinfo {author} {\bibfnamefont {R.~F.~L.}\
  \bibnamefont {{Holanda}}},\ }\href {\doibase
  10.1016/j.astropartphys.2016.09.001} {\bibfield  {journal} {\bibinfo
  {journal} {Astroparticle Physics}\ }\textbf {\bibinfo {volume} {84}},\
  \bibinfo {pages} {78} (\bibinfo {year} {2016})},\ \Eprint
  {http://arxiv.org/abs/1605.09415} {arXiv:1605.09415 [astro-ph.CO]}
  \BibitemShut {NoStop}%
\bibitem [{\citenamefont {{Li}}\ \emph {et~al.}(2011)\citenamefont {{Li}},
  \citenamefont {{Wu}},\ and\ \citenamefont {{Yu}}}]{2011ApJ...729L..14L}%
  \BibitemOpen
  \bibfield  {author} {\bibinfo {author} {\bibfnamefont {Z.}~\bibnamefont
  {{Li}}}, \bibinfo {author} {\bibfnamefont {P.}~\bibnamefont {{Wu}}}, \ and\
  \bibinfo {author} {\bibfnamefont {H.}~\bibnamefont {{Yu}}},\ }\href {\doibase
  10.1088/2041-8205/729/1/L14} {\bibfield  {journal} {\bibinfo  {journal}
  {ApJL}\ }\textbf {\bibinfo {volume} {729}},\ \bibinfo {eid} {L14} (\bibinfo
  {year} {2011})},\ \Eprint {http://arxiv.org/abs/1101.5255} {arXiv:1101.5255
  [astro-ph.CO]} \BibitemShut {NoStop}%
\bibitem [{\citenamefont {{Holanda}}\ \emph
  {et~al.}(2012{\natexlab{a}})\citenamefont {{Holanda}}, \citenamefont
  {{Gon{\c{c}}alves}},\ and\ \citenamefont {{Alcaniz}}}]{2012JCAP...06..022H}%
  \BibitemOpen
  \bibfield  {author} {\bibinfo {author} {\bibfnamefont {R.~F.~L.}\
  \bibnamefont {{Holanda}}}, \bibinfo {author} {\bibfnamefont {R.~S.}\
  \bibnamefont {{Gon{\c{c}}alves}}}, \ and\ \bibinfo {author} {\bibfnamefont
  {J.~S.}\ \bibnamefont {{Alcaniz}}},\ }\href {\doibase
  10.1088/1475-7516/2012/06/022} {\bibfield  {journal} {\bibinfo  {journal}
  {JCAP}\ }\textbf {\bibinfo {volume} {2012}},\ \bibinfo {eid} {022} (\bibinfo
  {year} {2012}{\natexlab{a}})},\ \Eprint {http://arxiv.org/abs/1201.2378}
  {arXiv:1201.2378 [astro-ph.CO]} \BibitemShut {NoStop}%
\bibitem [{\citenamefont {{Ellis}}\ \emph {et~al.}(2013)\citenamefont
  {{Ellis}}, \citenamefont {{Poltis}}, \citenamefont {{Uzan}},\ and\
  \citenamefont {{Weltman}}}]{2013PhRvD..87j3530E}%
  \BibitemOpen
  \bibfield  {author} {\bibinfo {author} {\bibfnamefont {G.~F.~R.}\
  \bibnamefont {{Ellis}}}, \bibinfo {author} {\bibfnamefont {R.}~\bibnamefont
  {{Poltis}}}, \bibinfo {author} {\bibfnamefont {J.-P.}\ \bibnamefont
  {{Uzan}}}, \ and\ \bibinfo {author} {\bibfnamefont {A.}~\bibnamefont
  {{Weltman}}},\ }\href {\doibase 10.1103/PhysRevD.87.103530} {\bibfield
  {journal} {\bibinfo  {journal} {PRD}\ }\textbf {\bibinfo {volume} {87}},\
  \bibinfo {eid} {103530} (\bibinfo {year} {2013})},\ \Eprint
  {http://arxiv.org/abs/1301.1312} {arXiv:1301.1312 [astro-ph.CO]} \BibitemShut
  {NoStop}%
\bibitem [{\citenamefont {{Wu}}\ \emph {et~al.}(2015)\citenamefont {{Wu}},
  \citenamefont {{Li}}, \citenamefont {{Liu}},\ and\ \citenamefont
  {{Yu}}}]{2015PhRvD..92b3520W}%
  \BibitemOpen
  \bibfield  {author} {\bibinfo {author} {\bibfnamefont {P.}~\bibnamefont
  {{Wu}}}, \bibinfo {author} {\bibfnamefont {Z.}~\bibnamefont {{Li}}}, \bibinfo
  {author} {\bibfnamefont {X.}~\bibnamefont {{Liu}}}, \ and\ \bibinfo {author}
  {\bibfnamefont {H.}~\bibnamefont {{Yu}}},\ }\href {\doibase
  10.1103/PhysRevD.92.023520} {\bibfield  {journal} {\bibinfo  {journal} {PRD}\
  }\textbf {\bibinfo {volume} {92}},\ \bibinfo {eid} {023520} (\bibinfo {year}
  {2015})}\BibitemShut {NoStop}%
\bibitem [{\citenamefont {{Yang}}\ \emph {et~al.}(2019)\citenamefont {{Yang}},
  \citenamefont {{Holanda}},\ and\ \citenamefont {{Hu}}}]{2019APh...108...57Y}%
  \BibitemOpen
  \bibfield  {author} {\bibinfo {author} {\bibfnamefont {T.}~\bibnamefont
  {{Yang}}}, \bibinfo {author} {\bibfnamefont {R.~F.~L.}\ \bibnamefont
  {{Holanda}}}, \ and\ \bibinfo {author} {\bibfnamefont {B.}~\bibnamefont
  {{Hu}}},\ }\href {\doibase 10.1016/j.astropartphys.2019.01.005} {\bibfield
  {journal} {\bibinfo  {journal} {Astroparticle Physics}\ }\textbf {\bibinfo
  {volume} {108}},\ \bibinfo {pages} {57} (\bibinfo {year} {2019})}\BibitemShut
  {NoStop}%
\bibitem [{\citenamefont {{Xu}}\ \emph {et~al.}(2020)\citenamefont {{Xu}},
  \citenamefont {{Zhang}},\ and\ \citenamefont
  {{Huang}}}]{2020EPJC...80..838X}%
  \BibitemOpen
  \bibfield  {author} {\bibinfo {author} {\bibfnamefont {B.}~\bibnamefont
  {{Xu}}}, \bibinfo {author} {\bibfnamefont {K.}~\bibnamefont {{Zhang}}}, \
  and\ \bibinfo {author} {\bibfnamefont {Q.}~\bibnamefont {{Huang}}},\ }\href
  {\doibase 10.1140/epjc/s10052-020-8426-4} {\bibfield  {journal} {\bibinfo
  {journal} {European Physical Journal C}\ }\textbf {\bibinfo {volume} {80}},\
  \bibinfo {eid} {838} (\bibinfo {year} {2020})}\BibitemShut {NoStop}%
\bibitem [{\citenamefont {{da Silva}}\ \emph {et~al.}(2020)\citenamefont {{da
  Silva}}, \citenamefont {{Holanda}},\ and\ \citenamefont
  {{Silva}}}]{2020PhRvD.102f3513D}%
  \BibitemOpen
  \bibfield  {author} {\bibinfo {author} {\bibfnamefont {W.~J.~C.}\
  \bibnamefont {{da Silva}}}, \bibinfo {author} {\bibfnamefont {R.~F.~L.}\
  \bibnamefont {{Holanda}}}, \ and\ \bibinfo {author} {\bibfnamefont
  {R.}~\bibnamefont {{Silva}}},\ }\href {\doibase 10.1103/PhysRevD.102.063513}
  {\bibfield  {journal} {\bibinfo  {journal} {PRD}\ }\textbf {\bibinfo {volume}
  {102}},\ \bibinfo {eid} {063513} (\bibinfo {year} {2020})},\ \Eprint
  {http://arxiv.org/abs/2005.04131} {arXiv:2005.04131 [astro-ph.CO]}
  \BibitemShut {NoStop}%
\bibitem [{\citenamefont {{Santos-da-Costa}}\ \emph {et~al.}(2015)\citenamefont
  {{Santos-da-Costa}}, \citenamefont {{Busti}},\ and\ \citenamefont
  {{Holanda}}}]{2015JCAP...10..061S}%
  \BibitemOpen
  \bibfield  {author} {\bibinfo {author} {\bibfnamefont {S.}~\bibnamefont
  {{Santos-da-Costa}}}, \bibinfo {author} {\bibfnamefont {V.~C.}\ \bibnamefont
  {{Busti}}}, \ and\ \bibinfo {author} {\bibfnamefont {R.~F.~L.}\ \bibnamefont
  {{Holanda}}},\ }\href {\doibase 10.1088/1475-7516/2015/10/061} {\bibfield
  {journal} {\bibinfo  {journal} {JCAP}\ }\textbf {\bibinfo {volume} {2015}},\
  \bibinfo {pages} {061} (\bibinfo {year} {2015})},\ \Eprint
  {http://arxiv.org/abs/1506.00145} {arXiv:1506.00145 [astro-ph.CO]}
  \BibitemShut {NoStop}%
\bibitem [{\citenamefont {{Liu}}\ \emph {et~al.}(2021)\citenamefont {{Liu}},
  \citenamefont {{Cao}}, \citenamefont {{Zhang}}, \citenamefont {{Gong}},
  \citenamefont {{Guo}},\ and\ \citenamefont {{Zheng}}}]{2021EPJC...81..903L}%
  \BibitemOpen
  \bibfield  {author} {\bibinfo {author} {\bibfnamefont {T.}~\bibnamefont
  {{Liu}}}, \bibinfo {author} {\bibfnamefont {S.}~\bibnamefont {{Cao}}},
  \bibinfo {author} {\bibfnamefont {S.}~\bibnamefont {{Zhang}}}, \bibinfo
  {author} {\bibfnamefont {X.}~\bibnamefont {{Gong}}}, \bibinfo {author}
  {\bibfnamefont {W.}~\bibnamefont {{Guo}}}, \ and\ \bibinfo {author}
  {\bibfnamefont {C.}~\bibnamefont {{Zheng}}},\ }\href {\doibase
  10.1140/epjc/s10052-021-09713-5} {\bibfield  {journal} {\bibinfo  {journal}
  {European Physical Journal C}\ }\textbf {\bibinfo {volume} {81}},\ \bibinfo
  {eid} {903} (\bibinfo {year} {2021})},\ \Eprint
  {http://arxiv.org/abs/2110.00927} {arXiv:2110.00927 [astro-ph.CO]}
  \BibitemShut {NoStop}%
\bibitem [{\citenamefont {Teixeira}\ \emph {et~al.}(2025)\citenamefont
  {Teixeira}, \citenamefont {Giar\`e}, \citenamefont {Hogg}, \citenamefont
  {Montandon}, \citenamefont {Poudou},\ and\ \citenamefont
  {Poulin}}]{Teixeira:2025czm}%
  \BibitemOpen
  \bibfield  {author} {\bibinfo {author} {\bibfnamefont {E.~M.}\ \bibnamefont
  {Teixeira}}, \bibinfo {author} {\bibfnamefont {W.}~\bibnamefont {Giar\`e}},
  \bibinfo {author} {\bibfnamefont {N.~B.}\ \bibnamefont {Hogg}}, \bibinfo
  {author} {\bibfnamefont {T.}~\bibnamefont {Montandon}}, \bibinfo {author}
  {\bibfnamefont {A.}~\bibnamefont {Poudou}}, \ and\ \bibinfo {author}
  {\bibfnamefont {V.}~\bibnamefont {Poulin}},\ }\href@noop {} {\  (\bibinfo
  {year} {2025})},\ \Eprint {http://arxiv.org/abs/2504.10464} {arXiv:2504.10464
  [astro-ph.CO]} \BibitemShut {NoStop}%
\bibitem [{\citenamefont {{Holanda}}(2012)}]{2012IJMPD..2150008H}%
  \BibitemOpen
  \bibfield  {author} {\bibinfo {author} {\bibfnamefont {R.~F.~L.}\
  \bibnamefont {{Holanda}}},\ }\href {\doibase 10.1142/S0218271812500083}
  {\bibfield  {journal} {\bibinfo  {journal} {International Journal of Modern
  Physics D}\ }\textbf {\bibinfo {volume} {21}},\ \bibinfo {eid}
  {1250008-1-1250008-10} (\bibinfo {year} {2012})},\ \Eprint
  {http://arxiv.org/abs/1202.2309} {arXiv:1202.2309 [astro-ph.CO]} \BibitemShut
  {NoStop}%
\bibitem [{\citenamefont {{Afroz}}\ and\ \citenamefont
  {{Mukherjee}}(2025)}]{2025arXiv250416868A}%
  \BibitemOpen
  \bibfield  {author} {\bibinfo {author} {\bibfnamefont {S.}~\bibnamefont
  {{Afroz}}}\ and\ \bibinfo {author} {\bibfnamefont {S.}~\bibnamefont
  {{Mukherjee}}},\ }\href {\doibase 10.48550/arXiv.2504.16868} {\bibfield
  {journal} {\bibinfo  {journal} {arXiv e-prints}\ ,\ \bibinfo {eid}
  {arXiv:2504.16868}} (\bibinfo {year} {2025})},\ \Eprint
  {http://arxiv.org/abs/2504.16868} {arXiv:2504.16868 [astro-ph.CO]}
  \BibitemShut {NoStop}%
\bibitem [{\citenamefont {{de Carvalho}}\ \emph {et~al.}(2018)\citenamefont
  {{de Carvalho}}, \citenamefont {{Bernui}}, \citenamefont {{Carvalho}},
  \citenamefont {{Novaes}},\ and\ \citenamefont {{Xavier}}}]{Edilson18}%
  \BibitemOpen
  \bibfield  {author} {\bibinfo {author} {\bibfnamefont {E.}~\bibnamefont {{de
  Carvalho}}}, \bibinfo {author} {\bibfnamefont {A.}~\bibnamefont {{Bernui}}},
  \bibinfo {author} {\bibfnamefont {G.~C.}\ \bibnamefont {{Carvalho}}},
  \bibinfo {author} {\bibfnamefont {C.~P.}\ \bibnamefont {{Novaes}}}, \ and\
  \bibinfo {author} {\bibfnamefont {H.~S.}\ \bibnamefont {{Xavier}}},\ }\href
  {\doibase 10.1088/1475-7516/2018/04/064} {\bibfield  {journal} {\bibinfo
  {journal} {JCAP}\ }\textbf {\bibinfo {volume} {4}},\ \bibinfo {eid} {064}
  (\bibinfo {year} {2018})},\ \Eprint {http://arxiv.org/abs/1709.00113}
  {arXiv:1709.00113 [astro-ph.CO]} \BibitemShut {NoStop}%
\bibitem [{\citenamefont {{DESI Collaboration}}(2025)}]{DESI2}%
  \BibitemOpen
  \bibfield  {author} {\bibinfo {author} {\bibnamefont {{DESI
  Collaboration}}},\ }\href {\doibase 10.48550/arXiv.2503.14738} {\bibfield
  {journal} {\bibinfo  {journal} {arXiv e-prints}\ ,\ \bibinfo {eid}
  {arXiv:2503.14738}} (\bibinfo {year} {2025})},\ \Eprint
  {http://arxiv.org/abs/2503.14738} {arXiv:2503.14738 [astro-ph.CO]}
  \BibitemShut {NoStop}%
\bibitem [{\citenamefont {{S{\'a}nchez}}\ \emph {et~al.}(2011)\citenamefont
  {{S{\'a}nchez}}, \citenamefont {{Carnero}}, \citenamefont
  {{Garc{\'\i}a-Bellido}}, \citenamefont {{Gazta{\~n}aga}}, \citenamefont {{de
  Simoni}}, \citenamefont {{Crocce}}, \citenamefont {{Cabr{\'e}}},
  \citenamefont {{Fosalba}},\ and\ \citenamefont {{Alonso}}}]{Sanchez11}%
  \BibitemOpen
  \bibfield  {author} {\bibinfo {author} {\bibfnamefont {E.}~\bibnamefont
  {{S{\'a}nchez}}}, \bibinfo {author} {\bibfnamefont {A.}~\bibnamefont
  {{Carnero}}}, \bibinfo {author} {\bibfnamefont {J.}~\bibnamefont
  {{Garc{\'\i}a-Bellido}}}, \bibinfo {author} {\bibfnamefont {E.}~\bibnamefont
  {{Gazta{\~n}aga}}}, \bibinfo {author} {\bibfnamefont {F.}~\bibnamefont {{de
  Simoni}}}, \bibinfo {author} {\bibfnamefont {M.}~\bibnamefont {{Crocce}}},
  \bibinfo {author} {\bibfnamefont {A.}~\bibnamefont {{Cabr{\'e}}}}, \bibinfo
  {author} {\bibfnamefont {P.}~\bibnamefont {{Fosalba}}}, \ and\ \bibinfo
  {author} {\bibfnamefont {D.}~\bibnamefont {{Alonso}}},\ }\href {\doibase
  10.1111/j.1365-2966.2010.17679.x} {\bibfield  {journal} {\bibinfo  {journal}
  {MNRAS}\ }\textbf {\bibinfo {volume} {411}},\ \bibinfo {pages} {277}
  (\bibinfo {year} {2011})},\ \Eprint {http://arxiv.org/abs/1006.3226}
  {arXiv:1006.3226 [astro-ph.CO]} \BibitemShut {NoStop}%
\bibitem [{\citenamefont {{Menote}}\ and\ \citenamefont
  {{Marra}}(2022)}]{Menote22}%
  \BibitemOpen
  \bibfield  {author} {\bibinfo {author} {\bibfnamefont {R.}~\bibnamefont
  {{Menote}}}\ and\ \bibinfo {author} {\bibfnamefont {V.}~\bibnamefont
  {{Marra}}},\ }\href {\doibase 10.1093/mnras/stac847} {\bibfield  {journal}
  {\bibinfo  {journal} {MNRAS}\ }\textbf {\bibinfo {volume} {513}},\ \bibinfo
  {pages} {1600} (\bibinfo {year} {2022})},\ \Eprint
  {http://arxiv.org/abs/2112.10000} {arXiv:2112.10000 [astro-ph.CO]}
  \BibitemShut {NoStop}%
\bibitem [{\citenamefont {{de Carvalho}}\ \emph {et~al.}(2021)\citenamefont
  {{de Carvalho}}, \citenamefont {{Bernui}}, \citenamefont {{Avila}},
  \citenamefont {{Novaes}},\ and\ \citenamefont
  {{Nogueira-Cavalcante}}}]{Edilson21}%
  \BibitemOpen
  \bibfield  {author} {\bibinfo {author} {\bibfnamefont {E.}~\bibnamefont {{de
  Carvalho}}}, \bibinfo {author} {\bibfnamefont {A.}~\bibnamefont {{Bernui}}},
  \bibinfo {author} {\bibfnamefont {F.}~\bibnamefont {{Avila}}}, \bibinfo
  {author} {\bibfnamefont {C.~P.}\ \bibnamefont {{Novaes}}}, \ and\ \bibinfo
  {author} {\bibfnamefont {J.~P.}\ \bibnamefont {{Nogueira-Cavalcante}}},\
  }\href {\doibase 10.1051/0004-6361/202039936} {\bibfield  {journal} {\bibinfo
   {journal} {aap}\ }\textbf {\bibinfo {volume} {649}},\ \bibinfo {eid} {A20}
  (\bibinfo {year} {2021})},\ \Eprint {http://arxiv.org/abs/2103.14121}
  {arXiv:2103.14121 [astro-ph.CO]} \BibitemShut {NoStop}%
\bibitem [{\citenamefont {{Planck Collaboration}}(2020)}]{Planck20}%
  \BibitemOpen
  \bibfield  {author} {\bibinfo {author} {\bibnamefont {{Planck
  Collaboration}}},\ }\href {\doibase 10.1051/0004-6361/201833910} {\bibfield
  {journal} {\bibinfo  {journal} {aap}\ }\textbf {\bibinfo {volume} {641}},\
  \bibinfo {eid} {A6} (\bibinfo {year} {2020})},\ \Eprint
  {http://arxiv.org/abs/1807.06209} {arXiv:1807.06209 [astro-ph.CO]}
  \BibitemShut {NoStop}%
\bibitem [{\citenamefont {Bonilla}\ \emph {et~al.}(2021)\citenamefont
  {Bonilla}, \citenamefont {Kumar},\ and\ \citenamefont
  {Nunes}}]{Bonilla:2020wbn}%
  \BibitemOpen
  \bibfield  {author} {\bibinfo {author} {\bibfnamefont {A.}~\bibnamefont
  {Bonilla}}, \bibinfo {author} {\bibfnamefont {S.}~\bibnamefont {Kumar}}, \
  and\ \bibinfo {author} {\bibfnamefont {R.~C.}\ \bibnamefont {Nunes}},\ }\href
  {\doibase 10.1140/epjc/s10052-021-08925-z} {\bibfield  {journal} {\bibinfo
  {journal} {Eur. Phys. J. C}\ }\textbf {\bibinfo {volume} {81}},\ \bibinfo
  {pages} {127} (\bibinfo {year} {2021})},\ \Eprint
  {http://arxiv.org/abs/2011.07140} {arXiv:2011.07140 [astro-ph.CO]}
  \BibitemShut {NoStop}%
\bibitem [{\citenamefont {{Sunyaev}}\ and\ \citenamefont
  {{Zeldovich}}(1972)}]{1972CoASP...4..173S}%
  \BibitemOpen
  \bibfield  {author} {\bibinfo {author} {\bibfnamefont {R.~A.}\ \bibnamefont
  {{Sunyaev}}}\ and\ \bibinfo {author} {\bibfnamefont {Y.~B.}\ \bibnamefont
  {{Zeldovich}}},\ }\href@noop {} {\bibfield  {journal} {\bibinfo  {journal}
  {Comments on Astrophysics and Space Physics}\ }\textbf {\bibinfo {volume}
  {4}},\ \bibinfo {pages} {173} (\bibinfo {year} {1972})}\BibitemShut {NoStop}%
\bibitem [{\citenamefont {{Birkinshaw}}(1999)}]{1999PhR...310...97B}%
  \BibitemOpen
  \bibfield  {author} {\bibinfo {author} {\bibfnamefont {M.}~\bibnamefont
  {{Birkinshaw}}},\ }\href {\doibase 10.1016/S0370-1573(98)00080-5} {\bibfield
  {journal} {\bibinfo  {journal} {Phys. Rep.}\ }\textbf {\bibinfo {volume}
  {310}},\ \bibinfo {pages} {97} (\bibinfo {year} {1999})},\ \Eprint
  {http://arxiv.org/abs/astro-ph/9808050} {arXiv:astro-ph/9808050 [astro-ph]}
  \BibitemShut {NoStop}%
\bibitem [{\citenamefont {{Sarazin}}(1986)}]{1988xrec.book.....S}%
  \BibitemOpen
  \bibfield  {author} {\bibinfo {author} {\bibfnamefont {C.~L.}\ \bibnamefont
  {{Sarazin}}},\ }\href {\doibase 10.1103/RevModPhys.58.1} {\bibfield
  {journal} {\bibinfo  {journal} {Reviews of Modern Physics}\ }\textbf
  {\bibinfo {volume} {58}},\ \bibinfo {pages} {1} (\bibinfo {year}
  {1986})}\BibitemShut {NoStop}%
\bibitem [{\citenamefont {{Reese}}\ \emph {et~al.}(2002)\citenamefont
  {{Reese}}, \citenamefont {{Carlstrom}}, \citenamefont {{Joy}}, \citenamefont
  {{Mohr}}, \citenamefont {{Grego}},\ and\ \citenamefont
  {{Holzapfel}}}]{2002ApJ...581...53R}%
  \BibitemOpen
  \bibfield  {author} {\bibinfo {author} {\bibfnamefont {E.~D.}\ \bibnamefont
  {{Reese}}}, \bibinfo {author} {\bibfnamefont {J.~E.}\ \bibnamefont
  {{Carlstrom}}}, \bibinfo {author} {\bibfnamefont {M.}~\bibnamefont {{Joy}}},
  \bibinfo {author} {\bibfnamefont {J.~J.}\ \bibnamefont {{Mohr}}}, \bibinfo
  {author} {\bibfnamefont {L.}~\bibnamefont {{Grego}}}, \ and\ \bibinfo
  {author} {\bibfnamefont {W.~L.}\ \bibnamefont {{Holzapfel}}},\ }\href
  {\doibase 10.1086/344137} {\bibfield  {journal} {\bibinfo  {journal} {apj}\
  }\textbf {\bibinfo {volume} {581}},\ \bibinfo {pages} {53} (\bibinfo {year}
  {2002})},\ \Eprint {http://arxiv.org/abs/astro-ph/0205350}
  {arXiv:astro-ph/0205350 [astro-ph]} \BibitemShut {NoStop}%
\bibitem [{\citenamefont {{Carlstrom}}\ \emph {et~al.}(2002)\citenamefont
  {{Carlstrom}}, \citenamefont {{Holder}},\ and\ \citenamefont
  {{Reese}}}]{2002ARA&A..40..643C}%
  \BibitemOpen
  \bibfield  {author} {\bibinfo {author} {\bibfnamefont {J.~E.}\ \bibnamefont
  {{Carlstrom}}}, \bibinfo {author} {\bibfnamefont {G.~P.}\ \bibnamefont
  {{Holder}}}, \ and\ \bibinfo {author} {\bibfnamefont {E.~D.}\ \bibnamefont
  {{Reese}}},\ }\href {\doibase 10.1146/annurev.astro.40.060401.093803}
  {\bibfield  {journal} {\bibinfo  {journal} {araa}\ }\textbf {\bibinfo
  {volume} {40}},\ \bibinfo {pages} {643} (\bibinfo {year} {2002})},\ \Eprint
  {http://arxiv.org/abs/astro-ph/0208192} {arXiv:astro-ph/0208192 [astro-ph]}
  \BibitemShut {NoStop}%
\bibitem [{\citenamefont {{Uzan}}\ \emph {et~al.}(2004)\citenamefont {{Uzan}},
  \citenamefont {{Aghanim}},\ and\ \citenamefont
  {{Mellier}}}]{2004PhRvD..70h3533U}%
  \BibitemOpen
  \bibfield  {author} {\bibinfo {author} {\bibfnamefont {J.-P.}\ \bibnamefont
  {{Uzan}}}, \bibinfo {author} {\bibfnamefont {N.}~\bibnamefont {{Aghanim}}}, \
  and\ \bibinfo {author} {\bibfnamefont {Y.}~\bibnamefont {{Mellier}}},\ }\href
  {\doibase 10.1103/PhysRevD.70.083533} {\bibfield  {journal} {\bibinfo
  {journal} {PRD}\ }\textbf {\bibinfo {volume} {70}},\ \bibinfo {eid} {083533}
  (\bibinfo {year} {2004})},\ \Eprint {http://arxiv.org/abs/astro-ph/0405620}
  {arXiv:astro-ph/0405620 [astro-ph]} \BibitemShut {NoStop}%
\bibitem [{\citenamefont {{Holanda}}\ \emph {et~al.}(2013)\citenamefont
  {{Holanda}}, \citenamefont {{Alcaniz}},\ and\ \citenamefont
  {{Carvalho}}}]{2013JCAP...06..033H}%
  \BibitemOpen
  \bibfield  {author} {\bibinfo {author} {\bibfnamefont {R.~F.~L.}\
  \bibnamefont {{Holanda}}}, \bibinfo {author} {\bibfnamefont {J.~S.}\
  \bibnamefont {{Alcaniz}}}, \ and\ \bibinfo {author} {\bibfnamefont {J.~C.}\
  \bibnamefont {{Carvalho}}},\ }\href {\doibase 10.1088/1475-7516/2013/06/033}
  {\bibfield  {journal} {\bibinfo  {journal} {JCAP}\ }\textbf {\bibinfo
  {volume} {2013}},\ \bibinfo {eid} {033} (\bibinfo {year} {2013})},\ \Eprint
  {http://arxiv.org/abs/1303.3307} {arXiv:1303.3307 [astro-ph.CO]} \BibitemShut
  {NoStop}%
\bibitem [{\citenamefont {{Holanda}}\ \emph
  {et~al.}(2012{\natexlab{b}})\citenamefont {{Holanda}}, \citenamefont
  {{Cunha}}, \citenamefont {{Marassi}},\ and\ \citenamefont
  {{Lima}}}]{2012JCAP...02..035H}%
  \BibitemOpen
  \bibfield  {author} {\bibinfo {author} {\bibfnamefont {R.~F.~L.}\
  \bibnamefont {{Holanda}}}, \bibinfo {author} {\bibfnamefont {J.~V.}\
  \bibnamefont {{Cunha}}}, \bibinfo {author} {\bibfnamefont {L.}~\bibnamefont
  {{Marassi}}}, \ and\ \bibinfo {author} {\bibfnamefont {J.~A.~S.}\
  \bibnamefont {{Lima}}},\ }\href {\doibase 10.1088/1475-7516/2012/02/035}
  {\bibfield  {journal} {\bibinfo  {journal} {JCAP}\ }\textbf {\bibinfo
  {volume} {2012}},\ \bibinfo {eid} {035} (\bibinfo {year}
  {2012}{\natexlab{b}})},\ \Eprint {http://arxiv.org/abs/1006.4200}
  {arXiv:1006.4200 [astro-ph.CO]} \BibitemShut {NoStop}%
\bibitem [{\citenamefont {{Cola{\c{c}}o}}\ \emph {et~al.}(2023)\citenamefont
  {{Cola{\c{c}}o}}, \citenamefont {{Ferreira}}, \citenamefont {{Holanda}},
  \citenamefont {{Gonzalez}},\ and\ \citenamefont
  {{Nunes}}}]{colaço2023hubble}%
  \BibitemOpen
  \bibfield  {author} {\bibinfo {author} {\bibfnamefont {L.~R.}\ \bibnamefont
  {{Cola{\c{c}}o}}}, \bibinfo {author} {\bibfnamefont {M.~S.}\ \bibnamefont
  {{Ferreira}}}, \bibinfo {author} {\bibfnamefont {R.~F.~L.}\ \bibnamefont
  {{Holanda}}}, \bibinfo {author} {\bibfnamefont {J.~E.}\ \bibnamefont
  {{Gonzalez}}}, \ and\ \bibinfo {author} {\bibfnamefont {R.~C.}\ \bibnamefont
  {{Nunes}}},\ }\href {\doibase 10.48550/arXiv.2310.18711} {\bibfield
  {journal} {\bibinfo  {journal} {arXiv e-prints}\ ,\ \bibinfo {eid}
  {arXiv:2310.18711}} (\bibinfo {year} {2023})},\ \Eprint
  {http://arxiv.org/abs/2310.18711} {arXiv:2310.18711 [astro-ph.CO]}
  \BibitemShut {NoStop}%
\bibitem [{\citenamefont {{De Filippis}}\ \emph {et~al.}(2005)\citenamefont
  {{De Filippis}}, \citenamefont {{Sereno}}, \citenamefont {{Bautz}},\ and\
  \citenamefont {{Longo}}}]{2005ApJ...625..108D}%
  \BibitemOpen
  \bibfield  {author} {\bibinfo {author} {\bibfnamefont {E.}~\bibnamefont {{De
  Filippis}}}, \bibinfo {author} {\bibfnamefont {M.}~\bibnamefont {{Sereno}}},
  \bibinfo {author} {\bibfnamefont {M.~W.}\ \bibnamefont {{Bautz}}}, \ and\
  \bibinfo {author} {\bibfnamefont {G.}~\bibnamefont {{Longo}}},\ }\href
  {\doibase 10.1086/429401} {\bibfield  {journal} {\bibinfo  {journal} {apj}\
  }\textbf {\bibinfo {volume} {625}},\ \bibinfo {pages} {108} (\bibinfo {year}
  {2005})},\ \Eprint {http://arxiv.org/abs/astro-ph/0502153}
  {arXiv:astro-ph/0502153 [astro-ph]} \BibitemShut {NoStop}%
\bibitem [{\citenamefont {{Bonamente}}\ \emph {et~al.}(2006)\citenamefont
  {{Bonamente}}, \citenamefont {{Joy}}, \citenamefont {{LaRoque}},
  \citenamefont {{Carlstrom}}, \citenamefont {{Reese}},\ and\ \citenamefont
  {{Dawson}}}]{Bonamente06}%
  \BibitemOpen
  \bibfield  {author} {\bibinfo {author} {\bibfnamefont {M.}~\bibnamefont
  {{Bonamente}}}, \bibinfo {author} {\bibfnamefont {M.~K.}\ \bibnamefont
  {{Joy}}}, \bibinfo {author} {\bibfnamefont {S.~J.}\ \bibnamefont
  {{LaRoque}}}, \bibinfo {author} {\bibfnamefont {J.~E.}\ \bibnamefont
  {{Carlstrom}}}, \bibinfo {author} {\bibfnamefont {E.~D.}\ \bibnamefont
  {{Reese}}}, \ and\ \bibinfo {author} {\bibfnamefont {K.~S.}\ \bibnamefont
  {{Dawson}}},\ }\href {\doibase 10.1086/505291} {\bibfield  {journal}
  {\bibinfo  {journal} {apj}\ }\textbf {\bibinfo {volume} {647}},\ \bibinfo
  {pages} {25} (\bibinfo {year} {2006})},\ \Eprint
  {http://arxiv.org/abs/astro-ph/0512349} {arXiv:astro-ph/0512349 [astro-ph]}
  \BibitemShut {NoStop}%
\bibitem [{\citenamefont {{Holanda}}\ \emph
  {et~al.}(2012{\natexlab{c}})\citenamefont {{Holanda}}, \citenamefont
  {{Lima}},\ and\ \citenamefont {{Ribeiro}}}]{2012A&A...538A.131H}%
  \BibitemOpen
  \bibfield  {author} {\bibinfo {author} {\bibfnamefont {R.~F.~L.}\
  \bibnamefont {{Holanda}}}, \bibinfo {author} {\bibfnamefont {J.~A.~S.}\
  \bibnamefont {{Lima}}}, \ and\ \bibinfo {author} {\bibfnamefont {M.~B.}\
  \bibnamefont {{Ribeiro}}},\ }\href {\doibase 10.1051/0004-6361/201118343}
  {\bibfield  {journal} {\bibinfo  {journal} {aap}\ }\textbf {\bibinfo {volume}
  {538}},\ \bibinfo {eid} {A131} (\bibinfo {year} {2012}{\natexlab{c}})},\
  \Eprint {http://arxiv.org/abs/1104.3753} {arXiv:1104.3753 [astro-ph.CO]}
  \BibitemShut {NoStop}%
\bibitem [{\citenamefont {Brout}\ \emph {et~al.}(2022)\citenamefont {Brout}
  \emph {et~al.}}]{Pantheon2022}%
  \BibitemOpen
  \bibfield  {author} {\bibinfo {author} {\bibfnamefont {D.}~\bibnamefont
  {Brout}} \emph {et~al.},\ }\href {\doibase 10.3847/1538-4357/ac8e04}
  {\bibfield  {journal} {\bibinfo  {journal} {Astrophys. J.}\ }\textbf
  {\bibinfo {volume} {938}},\ \bibinfo {pages} {110} (\bibinfo {year}
  {2022})},\ \Eprint {http://arxiv.org/abs/2202.04077} {arXiv:2202.04077
  [astro-ph.CO]} \BibitemShut {NoStop}%
\bibitem [{\citenamefont {{Risaliti}}\ and\ \citenamefont
  {{Lusso}}(2015)}]{Lusso15}%
  \BibitemOpen
  \bibfield  {author} {\bibinfo {author} {\bibfnamefont {G.}~\bibnamefont
  {{Risaliti}}}\ and\ \bibinfo {author} {\bibfnamefont {E.}~\bibnamefont
  {{Lusso}}},\ }\href {\doibase 10.1088/0004-637X/815/1/33} {\bibfield
  {journal} {\bibinfo  {journal} {apj}\ }\textbf {\bibinfo {volume} {815}},\
  \bibinfo {eid} {33} (\bibinfo {year} {2015})},\ \Eprint
  {http://arxiv.org/abs/1505.07118} {arXiv:1505.07118 [astro-ph.CO]}
  \BibitemShut {NoStop}%
\bibitem [{\citenamefont {{Risaliti}}\ and\ \citenamefont
  {{Lusso}}(2019)}]{Lusso19}%
  \BibitemOpen
  \bibfield  {author} {\bibinfo {author} {\bibfnamefont {G.}~\bibnamefont
  {{Risaliti}}}\ and\ \bibinfo {author} {\bibfnamefont {E.}~\bibnamefont
  {{Lusso}}},\ }\href {\doibase 10.1038/s41550-018-0657-z} {\bibfield
  {journal} {\bibinfo  {journal} {Nature Astronomy}\ }\textbf {\bibinfo
  {volume} {3}},\ \bibinfo {pages} {272} (\bibinfo {year} {2019})},\ \Eprint
  {http://arxiv.org/abs/1811.02590} {arXiv:1811.02590 [astro-ph.CO]}
  \BibitemShut {NoStop}%
\bibitem [{\citenamefont {{Lusso}}\ \emph {et~al.}(2020)\citenamefont
  {{Lusso}}, \citenamefont {{Risaliti}}, \citenamefont {{Nardini}},
  \citenamefont {{Bargiacchi}}, \citenamefont {{Benetti}}, \citenamefont
  {{Bisogni}}, \citenamefont {{Capozziello}}, \citenamefont {{Civano}},
  \citenamefont {{Eggleston}}, \citenamefont {{Elvis}}, \citenamefont
  {{Fabbiano}}, \citenamefont {{Gilli}}, \citenamefont {{Marconi}},
  \citenamefont {{Paolillo}}, \citenamefont {{Piedipalumbo}}, \citenamefont
  {{Salvestrini}}, \citenamefont {{Signorini}},\ and\ \citenamefont
  {{Vignali}}}]{Lusso20}%
  \BibitemOpen
  \bibfield  {author} {\bibinfo {author} {\bibfnamefont {E.}~\bibnamefont
  {{Lusso}}}, \bibinfo {author} {\bibfnamefont {G.}~\bibnamefont {{Risaliti}}},
  \bibinfo {author} {\bibfnamefont {E.}~\bibnamefont {{Nardini}}}, \bibinfo
  {author} {\bibfnamefont {G.}~\bibnamefont {{Bargiacchi}}}, \bibinfo {author}
  {\bibfnamefont {M.}~\bibnamefont {{Benetti}}}, \bibinfo {author}
  {\bibfnamefont {S.}~\bibnamefont {{Bisogni}}}, \bibinfo {author}
  {\bibfnamefont {S.}~\bibnamefont {{Capozziello}}}, \bibinfo {author}
  {\bibfnamefont {F.}~\bibnamefont {{Civano}}}, \bibinfo {author}
  {\bibfnamefont {L.}~\bibnamefont {{Eggleston}}}, \bibinfo {author}
  {\bibfnamefont {M.}~\bibnamefont {{Elvis}}}, \bibinfo {author} {\bibfnamefont
  {G.}~\bibnamefont {{Fabbiano}}}, \bibinfo {author} {\bibfnamefont
  {R.}~\bibnamefont {{Gilli}}}, \bibinfo {author} {\bibfnamefont
  {A.}~\bibnamefont {{Marconi}}}, \bibinfo {author} {\bibfnamefont
  {M.}~\bibnamefont {{Paolillo}}}, \bibinfo {author} {\bibfnamefont
  {E.}~\bibnamefont {{Piedipalumbo}}}, \bibinfo {author} {\bibfnamefont
  {F.}~\bibnamefont {{Salvestrini}}}, \bibinfo {author} {\bibfnamefont
  {M.}~\bibnamefont {{Signorini}}}, \ and\ \bibinfo {author} {\bibfnamefont
  {C.}~\bibnamefont {{Vignali}}},\ }\href {\doibase
  10.1051/0004-6361/202038899} {\bibfield  {journal} {\bibinfo  {journal}
  {aap}\ }\textbf {\bibinfo {volume} {642}},\ \bibinfo {eid} {A150} (\bibinfo
  {year} {2020})},\ \Eprint {http://arxiv.org/abs/2008.08586} {arXiv:2008.08586
  [astro-ph.GA]} \BibitemShut {NoStop}%
\bibitem [{\citenamefont {{Raffai}}\ \emph {et~al.}(2025)\citenamefont
  {{Raffai}}, \citenamefont {{Pataki}}, \citenamefont {{B{\"o}ttger}},
  \citenamefont {{Karsai}},\ and\ \citenamefont {{D{\'a}lya}}}]{Raffai25}%
  \BibitemOpen
  \bibfield  {author} {\bibinfo {author} {\bibfnamefont {P.}~\bibnamefont
  {{Raffai}}}, \bibinfo {author} {\bibfnamefont {A.}~\bibnamefont {{Pataki}}},
  \bibinfo {author} {\bibfnamefont {R.~L.}\ \bibnamefont {{B{\"o}ttger}}},
  \bibinfo {author} {\bibfnamefont {A.}~\bibnamefont {{Karsai}}}, \ and\
  \bibinfo {author} {\bibfnamefont {G.}~\bibnamefont {{D{\'a}lya}}},\ }\href
  {\doibase 10.3847/1538-4357/ada249} {\bibfield  {journal} {\bibinfo
  {journal} {apj}\ }\textbf {\bibinfo {volume} {979}},\ \bibinfo {eid} {51}
  (\bibinfo {year} {2025})},\ \Eprint {http://arxiv.org/abs/2412.15717}
  {arXiv:2412.15717 [astro-ph.CO]} \BibitemShut {NoStop}%
\bibitem [{\citenamefont {{Li}}\ \emph {et~al.}(2025)\citenamefont {{Li}},
  \citenamefont {{Keeley}},\ and\ \citenamefont {{Shafieloo}}}]{Li25}%
  \BibitemOpen
  \bibfield  {author} {\bibinfo {author} {\bibfnamefont {X.}~\bibnamefont
  {{Li}}}, \bibinfo {author} {\bibfnamefont {R.~E.}\ \bibnamefont {{Keeley}}},
  \ and\ \bibinfo {author} {\bibfnamefont {A.}~\bibnamefont {{Shafieloo}}},\
  }\href {\doibase 10.3847/1538-4357/adc2fe} {\bibfield  {journal} {\bibinfo
  {journal} {apj}\ }\textbf {\bibinfo {volume} {983}},\ \bibinfo {eid} {141}
  (\bibinfo {year} {2025})},\ \Eprint {http://arxiv.org/abs/2408.15547}
  {arXiv:2408.15547 [astro-ph.CO]} \BibitemShut {NoStop}%
\bibitem [{\citenamefont {{Seikel}}\ \emph {et~al.}(2012)\citenamefont
  {{Seikel}}, \citenamefont {{Clarkson}},\ and\ \citenamefont
  {{Smith}}}]{Seikel2012}%
  \BibitemOpen
  \bibfield  {author} {\bibinfo {author} {\bibfnamefont {M.}~\bibnamefont
  {{Seikel}}}, \bibinfo {author} {\bibfnamefont {C.}~\bibnamefont
  {{Clarkson}}}, \ and\ \bibinfo {author} {\bibfnamefont {M.}~\bibnamefont
  {{Smith}}},\ }\href {\doibase 10.1088/1475-7516/2012/06/036} {\bibfield
  {journal} {\bibinfo  {journal} {JCAP}\ }\textbf {\bibinfo {volume} {2012}},\
  \bibinfo {eid} {036} (\bibinfo {year} {2012})},\ \Eprint
  {http://arxiv.org/abs/1204.2832} {arXiv:1204.2832 [astro-ph.CO]} \BibitemShut
  {NoStop}%
\bibitem [{\citenamefont {{Seikel}}\ and\ \citenamefont
  {{Clarkson}}(2013)}]{Seikel13}%
  \BibitemOpen
  \bibfield  {author} {\bibinfo {author} {\bibfnamefont {M.}~\bibnamefont
  {{Seikel}}}\ and\ \bibinfo {author} {\bibfnamefont {C.}~\bibnamefont
  {{Clarkson}}},\ }\href {\doibase 10.48550/arXiv.1311.6678} {\bibfield
  {journal} {\bibinfo  {journal} {arXiv e-prints}\ ,\ \bibinfo {eid}
  {arXiv:1311.6678}} (\bibinfo {year} {2013})},\ \Eprint
  {http://arxiv.org/abs/1311.6678} {arXiv:1311.6678 [astro-ph.CO]} \BibitemShut
  {NoStop}%
\bibitem [{\citenamefont {{Yang}}\ \emph {et~al.}(2015)\citenamefont {{Yang}},
  \citenamefont {{Guo}},\ and\ \citenamefont {{Cai}}}]{Yang15}%
  \BibitemOpen
  \bibfield  {author} {\bibinfo {author} {\bibfnamefont {T.}~\bibnamefont
  {{Yang}}}, \bibinfo {author} {\bibfnamefont {Z.-K.}\ \bibnamefont {{Guo}}}, \
  and\ \bibinfo {author} {\bibfnamefont {R.-G.}\ \bibnamefont {{Cai}}},\ }\href
  {\doibase 10.1103/PhysRevD.91.123533} {\bibfield  {journal} {\bibinfo
  {journal} {PRD}\ }\textbf {\bibinfo {volume} {91}},\ \bibinfo {eid} {123533}
  (\bibinfo {year} {2015})},\ \Eprint {http://arxiv.org/abs/1505.04443}
  {arXiv:1505.04443 [astro-ph.CO]} \BibitemShut {NoStop}%
\bibitem [{\citenamefont {{G{\'o}mez-Valent}}\ and\ \citenamefont
  {{Amendola}}(2018)}]{Valente18}%
  \BibitemOpen
  \bibfield  {author} {\bibinfo {author} {\bibfnamefont {A.}~\bibnamefont
  {{G{\'o}mez-Valent}}}\ and\ \bibinfo {author} {\bibfnamefont
  {L.}~\bibnamefont {{Amendola}}},\ }\href {\doibase
  10.1088/1475-7516/2018/04/051} {\bibfield  {journal} {\bibinfo  {journal}
  {JCAP}\ }\textbf {\bibinfo {volume} {2018}},\ \bibinfo {eid} {051} (\bibinfo
  {year} {2018})},\ \Eprint {http://arxiv.org/abs/1802.01505} {arXiv:1802.01505
  [astro-ph.CO]} \BibitemShut {NoStop}%
\bibitem [{\citenamefont {{Zhang}}\ and\ \citenamefont {{Li}}(2018)}]{Zhang18}%
  \BibitemOpen
  \bibfield  {author} {\bibinfo {author} {\bibfnamefont {M.-J.}\ \bibnamefont
  {{Zhang}}}\ and\ \bibinfo {author} {\bibfnamefont {H.}~\bibnamefont {{Li}}},\
  }\href {\doibase 10.1140/epjc/s10052-018-5953-3} {\bibfield  {journal}
  {\bibinfo  {journal} {European Physical Journal C}\ }\textbf {\bibinfo
  {volume} {78}},\ \bibinfo {eid} {460} (\bibinfo {year} {2018})},\ \Eprint
  {http://arxiv.org/abs/1806.02981} {arXiv:1806.02981 [astro-ph.CO]}
  \BibitemShut {NoStop}%
\bibitem [{\citenamefont {Perenon}\ \emph {et~al.}(2022)\citenamefont
  {Perenon}, \citenamefont {Martinelli}, \citenamefont {Maartens},
  \citenamefont {Camera},\ and\ \citenamefont {Clarkson}}]{Perenon:2022fgw}%
  \BibitemOpen
  \bibfield  {author} {\bibinfo {author} {\bibfnamefont {L.}~\bibnamefont
  {Perenon}}, \bibinfo {author} {\bibfnamefont {M.}~\bibnamefont {Martinelli}},
  \bibinfo {author} {\bibfnamefont {R.}~\bibnamefont {Maartens}}, \bibinfo
  {author} {\bibfnamefont {S.}~\bibnamefont {Camera}}, \ and\ \bibinfo {author}
  {\bibfnamefont {C.}~\bibnamefont {Clarkson}},\ }\href {\doibase
  10.1016/j.dark.2022.101119} {\bibfield  {journal} {\bibinfo  {journal} {Phys.
  Dark Univ.}\ }\textbf {\bibinfo {volume} {37}},\ \bibinfo {pages} {101119}
  (\bibinfo {year} {2022})},\ \Eprint {http://arxiv.org/abs/2206.12375}
  {arXiv:2206.12375 [astro-ph.CO]} \BibitemShut {NoStop}%
\bibitem [{\citenamefont {Bonilla}\ \emph {et~al.}(2022)\citenamefont
  {Bonilla}, \citenamefont {Kumar}, \citenamefont {Nunes},\ and\ \citenamefont
  {Pan}}]{Bonilla2022}%
  \BibitemOpen
  \bibfield  {author} {\bibinfo {author} {\bibfnamefont {A.}~\bibnamefont
  {Bonilla}}, \bibinfo {author} {\bibfnamefont {S.}~\bibnamefont {Kumar}},
  \bibinfo {author} {\bibfnamefont {R.~C.}\ \bibnamefont {Nunes}}, \ and\
  \bibinfo {author} {\bibfnamefont {S.}~\bibnamefont {Pan}},\ }\href {\doibase
  10.1093/mnras/stac687} {\bibfield  {journal} {\bibinfo  {journal} {Mon. Not.
  Roy. Astron. Soc.}\ }\textbf {\bibinfo {volume} {512}},\ \bibinfo {pages}
  {4231} (\bibinfo {year} {2022})},\ \Eprint {http://arxiv.org/abs/2102.06149}
  {arXiv:2102.06149 [astro-ph.CO]} \BibitemShut {NoStop}%
\bibitem [{\citenamefont {Abedin}\ \emph {et~al.}(2025)\citenamefont {Abedin},
  \citenamefont {Wang}, \citenamefont {Ma},\ and\ \citenamefont
  {Pan}}]{Abedin:2025yru}%
  \BibitemOpen
  \bibfield  {author} {\bibinfo {author} {\bibfnamefont {M.}~\bibnamefont
  {Abedin}}, \bibinfo {author} {\bibfnamefont {G.-J.}\ \bibnamefont {Wang}},
  \bibinfo {author} {\bibfnamefont {Y.-Z.}\ \bibnamefont {Ma}}, \ and\ \bibinfo
  {author} {\bibfnamefont {S.}~\bibnamefont {Pan}},\ }\href@noop {} {\
  (\bibinfo {year} {2025})},\ \Eprint {http://arxiv.org/abs/2505.04336}
  {arXiv:2505.04336 [astro-ph.CO]} \BibitemShut {NoStop}%
\bibitem [{\citenamefont {{Jesus}}\ \emph {et~al.}(2020)\citenamefont
  {{Jesus}}, \citenamefont {{Valentim}}, \citenamefont {{Escobal}},\ and\
  \citenamefont {{Pereira}}}]{Jesus2019}%
  \BibitemOpen
  \bibfield  {author} {\bibinfo {author} {\bibfnamefont {J.~F.}\ \bibnamefont
  {{Jesus}}}, \bibinfo {author} {\bibfnamefont {R.}~\bibnamefont {{Valentim}}},
  \bibinfo {author} {\bibfnamefont {A.~A.}\ \bibnamefont {{Escobal}}}, \ and\
  \bibinfo {author} {\bibfnamefont {S.~H.}\ \bibnamefont {{Pereira}}},\ }\href
  {\doibase 10.1088/1475-7516/2020/04/053} {\bibfield  {journal} {\bibinfo
  {journal} {JCAP}\ }\textbf {\bibinfo {volume} {2020}},\ \bibinfo {eid} {053}
  (\bibinfo {year} {2020})},\ \Eprint {http://arxiv.org/abs/1909.00090}
  {arXiv:1909.00090 [astro-ph.CO]} \BibitemShut {NoStop}%
\bibitem [{\citenamefont {{Mukherjee}}\ and\ \citenamefont
  {{Banerjee}}(2020)}]{Mukherjee2020}%
  \BibitemOpen
  \bibfield  {author} {\bibinfo {author} {\bibfnamefont {P.}~\bibnamefont
  {{Mukherjee}}}\ and\ \bibinfo {author} {\bibfnamefont {N.}~\bibnamefont
  {{Banerjee}}},\ }\href {\doibase 10.48550/arXiv.2007.15941} {\bibfield
  {journal} {\bibinfo  {journal} {arXiv e-prints}\ ,\ \bibinfo {eid}
  {arXiv:2007.15941}} (\bibinfo {year} {2020})},\ \Eprint
  {http://arxiv.org/abs/2007.15941} {arXiv:2007.15941 [astro-ph.CO]}
  \BibitemShut {NoStop}%
\bibitem [{\citenamefont {Perenon}\ \emph {et~al.}(2021)\citenamefont
  {Perenon}, \citenamefont {Martinelli}, \citenamefont {Ili\'c}, \citenamefont
  {Maartens}, \citenamefont {Lochner},\ and\ \citenamefont
  {Clarkson}}]{Perenon21}%
  \BibitemOpen
  \bibfield  {author} {\bibinfo {author} {\bibfnamefont {L.}~\bibnamefont
  {Perenon}}, \bibinfo {author} {\bibfnamefont {M.}~\bibnamefont {Martinelli}},
  \bibinfo {author} {\bibfnamefont {S.}~\bibnamefont {Ili\'c}}, \bibinfo
  {author} {\bibfnamefont {R.}~\bibnamefont {Maartens}}, \bibinfo {author}
  {\bibfnamefont {M.}~\bibnamefont {Lochner}}, \ and\ \bibinfo {author}
  {\bibfnamefont {C.}~\bibnamefont {Clarkson}},\ }\href {\doibase
  10.1016/j.dark.2021.100898} {\bibfield  {journal} {\bibinfo  {journal} {Phys.
  Dark Univ.}\ }\textbf {\bibinfo {volume} {34}},\ \bibinfo {pages} {100898}
  (\bibinfo {year} {2021})},\ \Eprint {http://arxiv.org/abs/2105.01613}
  {arXiv:2105.01613 [astro-ph.CO]} \BibitemShut {NoStop}%
\bibitem [{\citenamefont {{Avila}}\ \emph
  {et~al.}(2022{\natexlab{a}})\citenamefont {{Avila}}, \citenamefont
  {{Bernui}}, \citenamefont {{Bonilla}},\ and\ \citenamefont
  {{Nunes}}}]{Avila22b}%
  \BibitemOpen
  \bibfield  {author} {\bibinfo {author} {\bibfnamefont {F.}~\bibnamefont
  {{Avila}}}, \bibinfo {author} {\bibfnamefont {A.}~\bibnamefont {{Bernui}}},
  \bibinfo {author} {\bibfnamefont {A.}~\bibnamefont {{Bonilla}}}, \ and\
  \bibinfo {author} {\bibfnamefont {R.~C.}\ \bibnamefont {{Nunes}}},\ }\href
  {\doibase 10.1140/epjc/s10052-022-10561-0} {\bibfield  {journal} {\bibinfo
  {journal} {European Physical Journal C}\ }\textbf {\bibinfo {volume} {82}},\
  \bibinfo {eid} {594} (\bibinfo {year} {2022}{\natexlab{a}})},\ \Eprint
  {http://arxiv.org/abs/2201.07829} {arXiv:2201.07829 [astro-ph.CO]}
  \BibitemShut {NoStop}%
\bibitem [{\citenamefont {Calder\'on}\ \emph {et~al.}(2023)\citenamefont
  {Calder\'on}, \citenamefont {L'Huillier}, \citenamefont {Polarski},
  \citenamefont {Shafieloo},\ and\ \citenamefont
  {Starobinsky}}]{Calderon2023msm}%
  \BibitemOpen
  \bibfield  {author} {\bibinfo {author} {\bibfnamefont {R.}~\bibnamefont
  {Calder\'on}}, \bibinfo {author} {\bibfnamefont {B.}~\bibnamefont
  {L'Huillier}}, \bibinfo {author} {\bibfnamefont {D.}~\bibnamefont
  {Polarski}}, \bibinfo {author} {\bibfnamefont {A.}~\bibnamefont {Shafieloo}},
  \ and\ \bibinfo {author} {\bibfnamefont {A.~A.}\ \bibnamefont
  {Starobinsky}},\ }\href {\doibase 10.1103/PhysRevD.108.023504} {\bibfield
  {journal} {\bibinfo  {journal} {Phys. Rev. D}\ }\textbf {\bibinfo {volume}
  {108}},\ \bibinfo {pages} {023504} (\bibinfo {year} {2023})},\ \Eprint
  {http://arxiv.org/abs/2301.00640} {arXiv:2301.00640 [astro-ph.CO]}
  \BibitemShut {NoStop}%
\bibitem [{\citenamefont {L'Huillier}\ \emph {et~al.}(2020)\citenamefont
  {L'Huillier}, \citenamefont {Shafieloo}, \citenamefont {Polarski},\ and\
  \citenamefont {Starobinsky}}]{LHuillier2019imn}%
  \BibitemOpen
  \bibfield  {author} {\bibinfo {author} {\bibfnamefont {B.}~\bibnamefont
  {L'Huillier}}, \bibinfo {author} {\bibfnamefont {A.}~\bibnamefont
  {Shafieloo}}, \bibinfo {author} {\bibfnamefont {D.}~\bibnamefont {Polarski}},
  \ and\ \bibinfo {author} {\bibfnamefont {A.~A.}\ \bibnamefont
  {Starobinsky}},\ }\href {\doibase 10.1093/mnras/staa633} {\bibfield
  {journal} {\bibinfo  {journal} {Mon. Not. Roy. Astron. Soc.}\ }\textbf
  {\bibinfo {volume} {494}},\ \bibinfo {pages} {819} (\bibinfo {year}
  {2020})},\ \Eprint {http://arxiv.org/abs/1906.05991} {arXiv:1906.05991
  [astro-ph.CO]} \BibitemShut {NoStop}%
\bibitem [{\citenamefont {{Oliveira}}\ \emph {et~al.}(2025)\citenamefont
  {{Oliveira}}, \citenamefont {{Avila}}, \citenamefont {{Franco}},\ and\
  \citenamefont {{Bernui}}}]{Oliveira25}%
  \BibitemOpen
  \bibfield  {author} {\bibinfo {author} {\bibfnamefont {F.}~\bibnamefont
  {{Oliveira}}}, \bibinfo {author} {\bibfnamefont {F.}~\bibnamefont {{Avila}}},
  \bibinfo {author} {\bibfnamefont {C.}~\bibnamefont {{Franco}}}, \ and\
  \bibinfo {author} {\bibfnamefont {A.}~\bibnamefont {{Bernui}}},\ }\href
  {\doibase 10.1016/j.dark.2025.101996} {\bibfield  {journal} {\bibinfo
  {journal} {Physics of the Dark Universe}\ }\textbf {\bibinfo {volume} {49}},\
  \bibinfo {eid} {101996} (\bibinfo {year} {2025})},\ \Eprint
  {http://arxiv.org/abs/2507.00779} {arXiv:2507.00779 [astro-ph.CO]}
  \BibitemShut {NoStop}%
\bibitem [{\citenamefont {{Avila}}\ \emph
  {et~al.}(2022{\natexlab{b}})\citenamefont {{Avila}}, \citenamefont
  {{Bernui}}, \citenamefont {{Nunes}}, \citenamefont {{de Carvalho}},\ and\
  \citenamefont {{Novaes}}}]{Avila22a}%
  \BibitemOpen
  \bibfield  {author} {\bibinfo {author} {\bibfnamefont {F.}~\bibnamefont
  {{Avila}}}, \bibinfo {author} {\bibfnamefont {A.}~\bibnamefont {{Bernui}}},
  \bibinfo {author} {\bibfnamefont {R.~C.}\ \bibnamefont {{Nunes}}}, \bibinfo
  {author} {\bibfnamefont {E.}~\bibnamefont {{de Carvalho}}}, \ and\ \bibinfo
  {author} {\bibfnamefont {C.~P.}\ \bibnamefont {{Novaes}}},\ }\href {\doibase
  10.1093/mnras/stab3122} {\bibfield  {journal} {\bibinfo  {journal} {MNRAS}\
  }\textbf {\bibinfo {volume} {509}},\ \bibinfo {pages} {2994} (\bibinfo {year}
  {2022}{\natexlab{b}})},\ \Eprint {http://arxiv.org/abs/2111.08541}
  {arXiv:2111.08541 [astro-ph.CO]} \BibitemShut {NoStop}%
\bibitem [{\citenamefont {{Yin}}\ and\ \citenamefont {{Wei}}(2019)}]{Yin19}%
  \BibitemOpen
  \bibfield  {author} {\bibinfo {author} {\bibfnamefont {Z.-Y.}\ \bibnamefont
  {{Yin}}}\ and\ \bibinfo {author} {\bibfnamefont {H.}~\bibnamefont {{Wei}}},\
  }\href {\doibase 10.1007/s11433-019-9373-0} {\bibfield  {journal} {\bibinfo
  {journal} {Science China Physics, Mechanics, and Astronomy}\ }\textbf
  {\bibinfo {volume} {62}},\ \bibinfo {eid} {999811} (\bibinfo {year}
  {2019})},\ \Eprint {http://arxiv.org/abs/1808.00377} {arXiv:1808.00377
  [astro-ph.CO]} \BibitemShut {NoStop}%
\bibitem [{\citenamefont {Mu}\ \emph {et~al.}(2023)\citenamefont {Mu},
  \citenamefont {Li},\ and\ \citenamefont {Xu}}]{Mu2023}%
  \BibitemOpen
  \bibfield  {author} {\bibinfo {author} {\bibfnamefont {Y.}~\bibnamefont
  {Mu}}, \bibinfo {author} {\bibfnamefont {E.-K.}\ \bibnamefont {Li}}, \ and\
  \bibinfo {author} {\bibfnamefont {L.}~\bibnamefont {Xu}},\ }\href {\doibase
  10.1088/1361-6382/acfb6c} {\bibfield  {journal} {\bibinfo  {journal}
  {Classical and Quantum Gravity}\ }\textbf {\bibinfo {volume} {40}},\ \bibinfo
  {pages} {225003} (\bibinfo {year} {2023})}\BibitemShut {NoStop}%
\bibitem [{\citenamefont {{Oliveira}}\ \emph {et~al.}(2024)\citenamefont
  {{Oliveira}}, \citenamefont {{Avila}}, \citenamefont {{Bernui}},
  \citenamefont {{Bonilla}},\ and\ \citenamefont {{Nunes}}}]{Oliveira23}%
  \BibitemOpen
  \bibfield  {author} {\bibinfo {author} {\bibfnamefont {F.}~\bibnamefont
  {{Oliveira}}}, \bibinfo {author} {\bibfnamefont {F.}~\bibnamefont {{Avila}}},
  \bibinfo {author} {\bibfnamefont {A.}~\bibnamefont {{Bernui}}}, \bibinfo
  {author} {\bibfnamefont {A.}~\bibnamefont {{Bonilla}}}, \ and\ \bibinfo
  {author} {\bibfnamefont {R.~C.}\ \bibnamefont {{Nunes}}},\ }\href {\doibase
  10.1140/epjc/s10052-024-12953-w} {\bibfield  {journal} {\bibinfo  {journal}
  {European Physical Journal C}\ }\textbf {\bibinfo {volume} {84}},\ \bibinfo
  {eid} {636} (\bibinfo {year} {2024})},\ \Eprint
  {http://arxiv.org/abs/2311.14216} {arXiv:2311.14216 [astro-ph.CO]}
  \BibitemShut {NoStop}%
\bibitem [{\citenamefont {Escamilla}\ \emph {et~al.}(2025)\citenamefont
  {Escamilla}, \citenamefont {Akarsu}, \citenamefont {Di~Valentino},
  \citenamefont {\"Oz\"ulker},\ and\ \citenamefont
  {Vazquez}}]{Escamilla:2025imi}%
  \BibitemOpen
  \bibfield  {author} {\bibinfo {author} {\bibfnamefont {L.~A.}\ \bibnamefont
  {Escamilla}}, \bibinfo {author} {\bibfnamefont {O.}~\bibnamefont {Akarsu}},
  \bibinfo {author} {\bibfnamefont {E.}~\bibnamefont {Di~Valentino}}, \bibinfo
  {author} {\bibfnamefont {E.}~\bibnamefont {\"Oz\"ulker}}, \ and\ \bibinfo
  {author} {\bibfnamefont {J.~A.}\ \bibnamefont {Vazquez}},\ }\href@noop {} {\
  (\bibinfo {year} {2025})},\ \Eprint {http://arxiv.org/abs/2503.12945}
  {arXiv:2503.12945 [astro-ph.CO]} \BibitemShut {NoStop}%
\bibitem [{\citenamefont {Sabogal}\ \emph {et~al.}(2024)\citenamefont
  {Sabogal}, \citenamefont {Akarsu}, \citenamefont {Bonilla}, \citenamefont
  {Di~Valentino},\ and\ \citenamefont {Nunes}}]{Sabogal:2024qxs}%
  \BibitemOpen
  \bibfield  {author} {\bibinfo {author} {\bibfnamefont {M.~A.}\ \bibnamefont
  {Sabogal}}, \bibinfo {author} {\bibfnamefont {O.}~\bibnamefont {Akarsu}},
  \bibinfo {author} {\bibfnamefont {A.}~\bibnamefont {Bonilla}}, \bibinfo
  {author} {\bibfnamefont {E.}~\bibnamefont {Di~Valentino}}, \ and\ \bibinfo
  {author} {\bibfnamefont {R.~C.}\ \bibnamefont {Nunes}},\ }\href {\doibase
  10.1140/epjc/s10052-024-13081-1} {\bibfield  {journal} {\bibinfo  {journal}
  {Eur. Phys. J. C}\ }\textbf {\bibinfo {volume} {84}},\ \bibinfo {pages} {703}
  (\bibinfo {year} {2024})},\ \Eprint {http://arxiv.org/abs/2407.04223}
  {arXiv:2407.04223 [astro-ph.CO]} \BibitemShut {NoStop}%
\bibitem [{\citenamefont {Gao}\ \emph {et~al.}(2025)\citenamefont {Gao},
  \citenamefont {Gao}, \citenamefont {Gong},\ and\ \citenamefont
  {Lu}}]{Gao:2025ozb}%
  \BibitemOpen
  \bibfield  {author} {\bibinfo {author} {\bibfnamefont {S.}~\bibnamefont
  {Gao}}, \bibinfo {author} {\bibfnamefont {Q.}~\bibnamefont {Gao}}, \bibinfo
  {author} {\bibfnamefont {Y.}~\bibnamefont {Gong}}, \ and\ \bibinfo {author}
  {\bibfnamefont {X.}~\bibnamefont {Lu}},\ }\href@noop {} {\  (\bibinfo {year}
  {2025})},\ \Eprint {http://arxiv.org/abs/2503.15943} {arXiv:2503.15943
  [astro-ph.CO]} \BibitemShut {NoStop}%
\bibitem [{\citenamefont {Jiang}\ \emph {et~al.}(2024)\citenamefont {Jiang},
  \citenamefont {Pedrotti}, \citenamefont {da~Costa},\ and\ \citenamefont
  {Vagnozzi}}]{Jiang:2024xnu}%
  \BibitemOpen
  \bibfield  {author} {\bibinfo {author} {\bibfnamefont {J.-Q.}\ \bibnamefont
  {Jiang}}, \bibinfo {author} {\bibfnamefont {D.}~\bibnamefont {Pedrotti}},
  \bibinfo {author} {\bibfnamefont {S.~S.}\ \bibnamefont {da~Costa}}, \ and\
  \bibinfo {author} {\bibfnamefont {S.}~\bibnamefont {Vagnozzi}},\ }\href
  {\doibase 10.1103/PhysRevD.110.123519} {\bibfield  {journal} {\bibinfo
  {journal} {Phys. Rev. D}\ }\textbf {\bibinfo {volume} {110}},\ \bibinfo
  {pages} {123519} (\bibinfo {year} {2024})},\ \Eprint
  {http://arxiv.org/abs/2408.02365} {arXiv:2408.02365 [astro-ph.CO]}
  \BibitemShut {NoStop}%
\bibitem [{\citenamefont {Dinda}\ and\ \citenamefont
  {Maartens}(2025)}]{Dinda:2024ktd}%
  \BibitemOpen
  \bibfield  {author} {\bibinfo {author} {\bibfnamefont {B.~R.}\ \bibnamefont
  {Dinda}}\ and\ \bibinfo {author} {\bibfnamefont {R.}~\bibnamefont
  {Maartens}},\ }\href {\doibase 10.1088/1475-7516/2025/01/120} {\bibfield
  {journal} {\bibinfo  {journal} {JCAP}\ }\textbf {\bibinfo {volume} {01}},\
  \bibinfo {pages} {120} (\bibinfo {year} {2025})},\ \Eprint
  {http://arxiv.org/abs/2407.17252} {arXiv:2407.17252 [astro-ph.CO]}
  \BibitemShut {NoStop}%
\bibitem [{\citenamefont {Dinda}\ \emph {et~al.}(2025)\citenamefont {Dinda},
  \citenamefont {Maartens}, \citenamefont {Saito},\ and\ \citenamefont
  {Clarkson}}]{Dinda:2025svh}%
  \BibitemOpen
  \bibfield  {author} {\bibinfo {author} {\bibfnamefont {B.~R.}\ \bibnamefont
  {Dinda}}, \bibinfo {author} {\bibfnamefont {R.}~\bibnamefont {Maartens}},
  \bibinfo {author} {\bibfnamefont {S.}~\bibnamefont {Saito}}, \ and\ \bibinfo
  {author} {\bibfnamefont {C.}~\bibnamefont {Clarkson}},\ }\href@noop {} {\
  (\bibinfo {year} {2025})},\ \Eprint {http://arxiv.org/abs/2504.09681}
  {arXiv:2504.09681 [astro-ph.CO]} \BibitemShut {NoStop}%
\bibitem [{\citenamefont {Yang}\ \emph
  {et~al.}(2025{\natexlab{a}})\citenamefont {Yang}, \citenamefont {Wang},
  \citenamefont {Li}, \citenamefont {Yuan}, \citenamefont {Ren}, \citenamefont
  {Saridakis},\ and\ \citenamefont {Cai}}]{Yang:2025kgc}%
  \BibitemOpen
  \bibfield  {author} {\bibinfo {author} {\bibfnamefont {Y.}~\bibnamefont
  {Yang}}, \bibinfo {author} {\bibfnamefont {Q.}~\bibnamefont {Wang}}, \bibinfo
  {author} {\bibfnamefont {C.}~\bibnamefont {Li}}, \bibinfo {author}
  {\bibfnamefont {P.}~\bibnamefont {Yuan}}, \bibinfo {author} {\bibfnamefont
  {X.}~\bibnamefont {Ren}}, \bibinfo {author} {\bibfnamefont {E.~N.}\
  \bibnamefont {Saridakis}}, \ and\ \bibinfo {author} {\bibfnamefont {Y.-F.}\
  \bibnamefont {Cai}},\ }\href@noop {} {\  (\bibinfo {year}
  {2025}{\natexlab{a}})},\ \Eprint {http://arxiv.org/abs/2501.18336}
  {arXiv:2501.18336 [astro-ph.CO]} \BibitemShut {NoStop}%
\bibitem [{\citenamefont {{Dinda}}\ and\ \citenamefont
  {{Banerjee}}(2023)}]{Dinda2023}%
  \BibitemOpen
  \bibfield  {author} {\bibinfo {author} {\bibfnamefont {B.~R.}\ \bibnamefont
  {{Dinda}}}\ and\ \bibinfo {author} {\bibfnamefont {N.}~\bibnamefont
  {{Banerjee}}},\ }\href {\doibase 10.1103/PhysRevD.107.063513} {\bibfield
  {journal} {\bibinfo  {journal} {PRD}\ }\textbf {\bibinfo {volume} {107}},\
  \bibinfo {eid} {063513} (\bibinfo {year} {2023})},\ \Eprint
  {http://arxiv.org/abs/2208.14740} {arXiv:2208.14740 [astro-ph.CO]}
  \BibitemShut {NoStop}%
\bibitem [{\citenamefont {{Ruiz-Zapatero}}\ \emph {et~al.}(2022)\citenamefont
  {{Ruiz-Zapatero}}, \citenamefont {{Garc{\'\i}a-Garc{\'\i}a}}, \citenamefont
  {{Alonso}}, \citenamefont {{Ferreira}},\ and\ \citenamefont
  {{Grumitt}}}]{RuizZapatero2022}%
  \BibitemOpen
  \bibfield  {author} {\bibinfo {author} {\bibfnamefont {J.}~\bibnamefont
  {{Ruiz-Zapatero}}}, \bibinfo {author} {\bibfnamefont {C.}~\bibnamefont
  {{Garc{\'\i}a-Garc{\'\i}a}}}, \bibinfo {author} {\bibfnamefont
  {D.}~\bibnamefont {{Alonso}}}, \bibinfo {author} {\bibfnamefont {P.~G.}\
  \bibnamefont {{Ferreira}}}, \ and\ \bibinfo {author} {\bibfnamefont
  {R.~D.~P.}\ \bibnamefont {{Grumitt}}},\ }\href {\doibase
  10.1093/mnras/stac431} {\bibfield  {journal} {\bibinfo  {journal} {MNRAS}\
  }\textbf {\bibinfo {volume} {512}},\ \bibinfo {pages} {1967} (\bibinfo {year}
  {2022})},\ \Eprint {http://arxiv.org/abs/2201.07025} {arXiv:2201.07025
  [astro-ph.CO]} \BibitemShut {NoStop}%
\bibitem [{\citenamefont {{Escamilla}}\ \emph {et~al.}(2023)\citenamefont
  {{Escamilla}}, \citenamefont {{Akarsu}}, \citenamefont {{Di Valentino}},\
  and\ \citenamefont {{Vazquez}}}]{Escamilla2023modelindependent}%
  \BibitemOpen
  \bibfield  {author} {\bibinfo {author} {\bibfnamefont {L.~A.}\ \bibnamefont
  {{Escamilla}}}, \bibinfo {author} {\bibfnamefont {{\"O}.}~\bibnamefont
  {{Akarsu}}}, \bibinfo {author} {\bibfnamefont {E.}~\bibnamefont {{Di
  Valentino}}}, \ and\ \bibinfo {author} {\bibfnamefont {J.~A.}\ \bibnamefont
  {{Vazquez}}},\ }\href {\doibase 10.1088/1475-7516/2023/11/051} {\bibfield
  {journal} {\bibinfo  {journal} {JCAP}\ }\textbf {\bibinfo {volume} {2023}},\
  \bibinfo {eid} {051} (\bibinfo {year} {2023})},\ \Eprint
  {http://arxiv.org/abs/2305.16290} {arXiv:2305.16290 [astro-ph.CO]}
  \BibitemShut {NoStop}%
\bibitem [{\citenamefont {{Sun}}\ \emph {et~al.}(2021)\citenamefont {{Sun}},
  \citenamefont {{Jiao}},\ and\ \citenamefont {{Zhang}}}]{Sun2021}%
  \BibitemOpen
  \bibfield  {author} {\bibinfo {author} {\bibfnamefont {W.}~\bibnamefont
  {{Sun}}}, \bibinfo {author} {\bibfnamefont {K.}~\bibnamefont {{Jiao}}}, \
  and\ \bibinfo {author} {\bibfnamefont {T.-J.}\ \bibnamefont {{Zhang}}},\
  }\href {\doibase 10.3847/1538-4357/ac05b8} {\bibfield  {journal} {\bibinfo
  {journal} {apj}\ }\textbf {\bibinfo {volume} {915}},\ \bibinfo {eid} {123}
  (\bibinfo {year} {2021})},\ \Eprint {http://arxiv.org/abs/2105.12618}
  {arXiv:2105.12618 [astro-ph.CO]} \BibitemShut {NoStop}%
\bibitem [{\citenamefont {\'O~Colg\'ain}\ and\ \citenamefont
  {Sheikh-Jabbari}(2021)}]{OColgain2021}%
  \BibitemOpen
  \bibfield  {author} {\bibinfo {author} {\bibfnamefont {E.}~\bibnamefont
  {\'O~Colg\'ain}}\ and\ \bibinfo {author} {\bibfnamefont {M.~M.}\ \bibnamefont
  {Sheikh-Jabbari}},\ }\href {\doibase 10.1140/epjc/s10052-021-09708-2}
  {\bibfield  {journal} {\bibinfo  {journal} {Eur. Phys. J. C}\ }\textbf
  {\bibinfo {volume} {81}},\ \bibinfo {pages} {892} (\bibinfo {year} {2021})},\
  \Eprint {http://arxiv.org/abs/2101.08565} {arXiv:2101.08565 [astro-ph.CO]}
  \BibitemShut {NoStop}%
\bibitem [{\citenamefont {{Ahlstr{\"o}m Kjerrgren}}\ and\ \citenamefont
  {{M{\"o}rtsell}}(2023)}]{Kjerrgren2021}%
  \BibitemOpen
  \bibfield  {author} {\bibinfo {author} {\bibfnamefont {A.}~\bibnamefont
  {{Ahlstr{\"o}m Kjerrgren}}}\ and\ \bibinfo {author} {\bibfnamefont
  {E.}~\bibnamefont {{M{\"o}rtsell}}},\ }\href {\doibase
  10.1093/mnras/stac1978} {\bibfield  {journal} {\bibinfo  {journal} {MNRAS}\
  }\textbf {\bibinfo {volume} {518}},\ \bibinfo {pages} {585} (\bibinfo {year}
  {2023})},\ \Eprint {http://arxiv.org/abs/2106.11317} {arXiv:2106.11317
  [astro-ph.CO]} \BibitemShut {NoStop}%
\bibitem [{\citenamefont {{Renzi}}\ and\ \citenamefont
  {{Silvestri}}(2020)}]{Renzi2020}%
  \BibitemOpen
  \bibfield  {author} {\bibinfo {author} {\bibfnamefont {F.}~\bibnamefont
  {{Renzi}}}\ and\ \bibinfo {author} {\bibfnamefont {A.}~\bibnamefont
  {{Silvestri}}},\ }\href {\doibase 10.48550/arXiv.2011.10559} {\bibfield
  {journal} {\bibinfo  {journal} {arXiv e-prints}\ ,\ \bibinfo {eid}
  {arXiv:2011.10559}} (\bibinfo {year} {2020})},\ \Eprint
  {http://arxiv.org/abs/2011.10559} {arXiv:2011.10559 [astro-ph.CO]}
  \BibitemShut {NoStop}%
\bibitem [{\citenamefont {Calder\'on}\ \emph {et~al.}(2022)\citenamefont
  {Calder\'on}, \citenamefont {L'Huillier}, \citenamefont {Polarski},
  \citenamefont {Shafieloo},\ and\ \citenamefont
  {Starobinsky}}]{Calderon2022cfj}%
  \BibitemOpen
  \bibfield  {author} {\bibinfo {author} {\bibfnamefont {R.}~\bibnamefont
  {Calder\'on}}, \bibinfo {author} {\bibfnamefont {B.}~\bibnamefont
  {L'Huillier}}, \bibinfo {author} {\bibfnamefont {D.}~\bibnamefont
  {Polarski}}, \bibinfo {author} {\bibfnamefont {A.}~\bibnamefont {Shafieloo}},
  \ and\ \bibinfo {author} {\bibfnamefont {A.~A.}\ \bibnamefont
  {Starobinsky}},\ }\href {\doibase 10.1103/PhysRevD.106.083513} {\bibfield
  {journal} {\bibinfo  {journal} {Phys. Rev. D}\ }\textbf {\bibinfo {volume}
  {106}},\ \bibinfo {pages} {083513} (\bibinfo {year} {2022})},\ \Eprint
  {http://arxiv.org/abs/2206.13820} {arXiv:2206.13820 [astro-ph.CO]}
  \BibitemShut {NoStop}%
\bibitem [{\citenamefont {{Dinda}}(2024)}]{Dinda2023xqx}%
  \BibitemOpen
  \bibfield  {author} {\bibinfo {author} {\bibfnamefont {B.~R.}\ \bibnamefont
  {{Dinda}}},\ }\href {\doibase 10.1140/epjc/s10052-024-12774-x} {\bibfield
  {journal} {\bibinfo  {journal} {European Physical Journal C}\ }\textbf
  {\bibinfo {volume} {84}},\ \bibinfo {eid} {402} (\bibinfo {year} {2024})},\
  \Eprint {http://arxiv.org/abs/2311.13498} {arXiv:2311.13498 [astro-ph.CO]}
  \BibitemShut {NoStop}%
\bibitem [{\citenamefont {Escamilla}\ \emph {et~al.}(2024)\citenamefont
  {Escamilla}, \citenamefont {\"Oz\"ulker}, \citenamefont {Akarsu},
  \citenamefont {Di~Valentino},\ and\ \citenamefont
  {V\'azquez}}]{Escamilla:2024ahl}%
  \BibitemOpen
  \bibfield  {author} {\bibinfo {author} {\bibfnamefont {L.~A.}\ \bibnamefont
  {Escamilla}}, \bibinfo {author} {\bibfnamefont {E.}~\bibnamefont
  {\"Oz\"ulker}}, \bibinfo {author} {\bibfnamefont {O.}~\bibnamefont {Akarsu}},
  \bibinfo {author} {\bibfnamefont {E.}~\bibnamefont {Di~Valentino}}, \ and\
  \bibinfo {author} {\bibfnamefont {J.~A.}\ \bibnamefont {V\'azquez}},\
  }\href@noop {} {\  (\bibinfo {year} {2024})},\ \Eprint
  {http://arxiv.org/abs/2408.12516} {arXiv:2408.12516 [astro-ph.CO]}
  \BibitemShut {NoStop}%
\bibitem [{\citenamefont {Vel\'azquez}\ \emph {et~al.}(2024)\citenamefont
  {Vel\'azquez}, \citenamefont {Escamilla}, \citenamefont {Mukherjee},\ and\
  \citenamefont {V\'azquez}}]{Velazquez:2024aya}%
  \BibitemOpen
  \bibfield  {author} {\bibinfo {author} {\bibfnamefont {J.~d.~J.}\
  \bibnamefont {Vel\'azquez}}, \bibinfo {author} {\bibfnamefont {L.~A.}\
  \bibnamefont {Escamilla}}, \bibinfo {author} {\bibfnamefont {P.}~\bibnamefont
  {Mukherjee}}, \ and\ \bibinfo {author} {\bibfnamefont {J.~A.}\ \bibnamefont
  {V\'azquez}},\ }\href {\doibase 10.3390/universe10120464} {\bibfield
  {journal} {\bibinfo  {journal} {Universe}\ }\textbf {\bibinfo {volume}
  {10}},\ \bibinfo {pages} {464} (\bibinfo {year} {2024})},\ \Eprint
  {http://arxiv.org/abs/2410.02061} {arXiv:2410.02061 [astro-ph.CO]}
  \BibitemShut {NoStop}%
\bibitem [{\citenamefont {G\'omez-Vargas}\ \emph {et~al.}(2023)\citenamefont
  {G\'omez-Vargas}, \citenamefont {Esquivel}, \citenamefont
  {Garc\'\i{}a-Salcedo},\ and\ \citenamefont
  {V\'azquez}}]{Gomez-Vargas:2021zyl}%
  \BibitemOpen
  \bibfield  {author} {\bibinfo {author} {\bibfnamefont {I.}~\bibnamefont
  {G\'omez-Vargas}}, \bibinfo {author} {\bibfnamefont {R.~M.}\ \bibnamefont
  {Esquivel}}, \bibinfo {author} {\bibfnamefont {R.}~\bibnamefont
  {Garc\'\i{}a-Salcedo}}, \ and\ \bibinfo {author} {\bibfnamefont {J.~A.}\
  \bibnamefont {V\'azquez}},\ }\href {\doibase 10.1140/epjc/s10052-023-11435-9}
  {\bibfield  {journal} {\bibinfo  {journal} {Eur. Phys. J. C}\ }\textbf
  {\bibinfo {volume} {83}},\ \bibinfo {pages} {304} (\bibinfo {year} {2023})},\
  \Eprint {http://arxiv.org/abs/2104.00595} {arXiv:2104.00595 [astro-ph.CO]}
  \BibitemShut {NoStop}%
\bibitem [{\citenamefont {Keeley}\ \emph {et~al.}(2021)\citenamefont {Keeley},
  \citenamefont {Shafieloo}, \citenamefont {Zhao}, \citenamefont {Vazquez},\
  and\ \citenamefont {Koo}}]{Keeley:2020aym}%
  \BibitemOpen
  \bibfield  {author} {\bibinfo {author} {\bibfnamefont {R.~E.}\ \bibnamefont
  {Keeley}}, \bibinfo {author} {\bibfnamefont {A.}~\bibnamefont {Shafieloo}},
  \bibinfo {author} {\bibfnamefont {G.-B.}\ \bibnamefont {Zhao}}, \bibinfo
  {author} {\bibfnamefont {J.~A.}\ \bibnamefont {Vazquez}}, \ and\ \bibinfo
  {author} {\bibfnamefont {H.}~\bibnamefont {Koo}},\ }\href {\doibase
  10.3847/1538-3881/abdd2a} {\bibfield  {journal} {\bibinfo  {journal} {Astron.
  J.}\ }\textbf {\bibinfo {volume} {161}},\ \bibinfo {pages} {151} (\bibinfo
  {year} {2021})},\ \Eprint {http://arxiv.org/abs/2010.03234} {arXiv:2010.03234
  [astro-ph.CO]} \BibitemShut {NoStop}%
\bibitem [{\citenamefont {Zheng}\ \emph {et~al.}(2025)\citenamefont {Zheng},
  \citenamefont {Qiang}, \citenamefont {You},\ and\ \citenamefont
  {Kumar}}]{Zheng:2025cgq}%
  \BibitemOpen
  \bibfield  {author} {\bibinfo {author} {\bibfnamefont {J.}~\bibnamefont
  {Zheng}}, \bibinfo {author} {\bibfnamefont {D.-C.}\ \bibnamefont {Qiang}},
  \bibinfo {author} {\bibfnamefont {Z.-Q.}\ \bibnamefont {You}}, \ and\
  \bibinfo {author} {\bibfnamefont {D.}~\bibnamefont {Kumar}},\ }\href@noop {}
  {\  (\bibinfo {year} {2025})},\ \Eprint {http://arxiv.org/abs/2507.17113}
  {arXiv:2507.17113 [astro-ph.CO]} \BibitemShut {NoStop}%
\bibitem [{\citenamefont {Li}\ \emph {et~al.}(2025)\citenamefont {Li},
  \citenamefont {Du}, \citenamefont {Wu}, \citenamefont {Qi}, \citenamefont
  {Zhang},\ and\ \citenamefont {Zhang}}]{Li:2025htp}%
  \BibitemOpen
  \bibfield  {author} {\bibinfo {author} {\bibfnamefont {T.-N.}\ \bibnamefont
  {Li}}, \bibinfo {author} {\bibfnamefont {G.-H.}\ \bibnamefont {Du}}, \bibinfo
  {author} {\bibfnamefont {P.-J.}\ \bibnamefont {Wu}}, \bibinfo {author}
  {\bibfnamefont {J.-Z.}\ \bibnamefont {Qi}}, \bibinfo {author} {\bibfnamefont
  {J.-F.}\ \bibnamefont {Zhang}}, \ and\ \bibinfo {author} {\bibfnamefont
  {X.}~\bibnamefont {Zhang}},\ }\href@noop {} {\  (\bibinfo {year} {2025})},\
  \Eprint {http://arxiv.org/abs/2507.13811} {arXiv:2507.13811 [astro-ph.CO]}
  \BibitemShut {NoStop}%
\bibitem [{\citenamefont {Yang}\ \emph
  {et~al.}(2025{\natexlab{b}})\citenamefont {Yang}, \citenamefont {Fu},
  \citenamefont {Xu}, \citenamefont {Zhang}, \citenamefont {Huang},\ and\
  \citenamefont {Yang}}]{Yang:2025qdg}%
  \BibitemOpen
  \bibfield  {author} {\bibinfo {author} {\bibfnamefont {F.}~\bibnamefont
  {Yang}}, \bibinfo {author} {\bibfnamefont {X.}~\bibnamefont {Fu}}, \bibinfo
  {author} {\bibfnamefont {B.}~\bibnamefont {Xu}}, \bibinfo {author}
  {\bibfnamefont {K.}~\bibnamefont {Zhang}}, \bibinfo {author} {\bibfnamefont
  {Y.}~\bibnamefont {Huang}}, \ and\ \bibinfo {author} {\bibfnamefont
  {Y.}~\bibnamefont {Yang}},\ }\href {\doibase 10.1140/epjc/s10052-025-13892-w}
  {\bibfield  {journal} {\bibinfo  {journal} {Eur. Phys. J. C}\ }\textbf
  {\bibinfo {volume} {85}},\ \bibinfo {pages} {186} (\bibinfo {year}
  {2025}{\natexlab{b}})},\ \Eprint {http://arxiv.org/abs/2502.05417}
  {arXiv:2502.05417 [astro-ph.CO]} \BibitemShut {NoStop}%
\bibitem [{\citenamefont {Cosmai}\ \emph {et~al.}(2013)\citenamefont {Cosmai},
  \citenamefont {Fanizza}, \citenamefont {Gasperini},\ and\ \citenamefont
  {Tedesco}}]{Cosmai:2013iga}%
  \BibitemOpen
  \bibfield  {author} {\bibinfo {author} {\bibfnamefont {L.}~\bibnamefont
  {Cosmai}}, \bibinfo {author} {\bibfnamefont {G.}~\bibnamefont {Fanizza}},
  \bibinfo {author} {\bibfnamefont {M.}~\bibnamefont {Gasperini}}, \ and\
  \bibinfo {author} {\bibfnamefont {L.}~\bibnamefont {Tedesco}},\ }\href
  {\doibase 10.1088/0264-9381/30/9/095011} {\bibfield  {journal} {\bibinfo
  {journal} {Class. Quant. Grav.}\ }\textbf {\bibinfo {volume} {30}},\ \bibinfo
  {pages} {095011} (\bibinfo {year} {2013})},\ \Eprint
  {http://arxiv.org/abs/1303.5484} {arXiv:1303.5484 [gr-qc]} \BibitemShut
  {NoStop}%
\bibitem [{\citenamefont {Wang}\ \emph {et~al.}(2024)\citenamefont {Wang},
  \citenamefont {Fu}, \citenamefont {Xu}, \citenamefont {Huang}, \citenamefont
  {Yang},\ and\ \citenamefont {Lu}}]{Wang:2024rxm}%
  \BibitemOpen
  \bibfield  {author} {\bibinfo {author} {\bibfnamefont {M.}~\bibnamefont
  {Wang}}, \bibinfo {author} {\bibfnamefont {X.}~\bibnamefont {Fu}}, \bibinfo
  {author} {\bibfnamefont {B.}~\bibnamefont {Xu}}, \bibinfo {author}
  {\bibfnamefont {Y.}~\bibnamefont {Huang}}, \bibinfo {author} {\bibfnamefont
  {Y.}~\bibnamefont {Yang}}, \ and\ \bibinfo {author} {\bibfnamefont
  {Z.}~\bibnamefont {Lu}},\ }\href {\doibase 10.1140/epjc/s10052-024-13049-1}
  {\bibfield  {journal} {\bibinfo  {journal} {Eur. Phys. J. C}\ }\textbf
  {\bibinfo {volume} {84}},\ \bibinfo {pages} {702} (\bibinfo {year} {2024})},\
  \Eprint {http://arxiv.org/abs/2407.12250} {arXiv:2407.12250 [astro-ph.CO]}
  \BibitemShut {NoStop}%
\bibitem [{\citenamefont {{Hwang}}\ \emph {et~al.}(2023)\citenamefont
  {{Hwang}}, \citenamefont {{L'Huillier}}, \citenamefont {{Keeley}},
  \citenamefont {{Jee}},\ and\ \citenamefont {{Shafieloo}}}]{Hwang23}%
  \BibitemOpen
  \bibfield  {author} {\bibinfo {author} {\bibfnamefont {S.-g.}\ \bibnamefont
  {{Hwang}}}, \bibinfo {author} {\bibfnamefont {B.}~\bibnamefont
  {{L'Huillier}}}, \bibinfo {author} {\bibfnamefont {R.~E.}\ \bibnamefont
  {{Keeley}}}, \bibinfo {author} {\bibfnamefont {M.~J.}\ \bibnamefont {{Jee}}},
  \ and\ \bibinfo {author} {\bibfnamefont {A.}~\bibnamefont {{Shafieloo}}},\
  }\href {\doibase 10.1088/1475-7516/2023/02/014} {\bibfield  {journal}
  {\bibinfo  {journal} {JCAP}\ }\textbf {\bibinfo {volume} {2023}},\ \bibinfo
  {eid} {014} (\bibinfo {year} {2023})},\ \Eprint
  {http://arxiv.org/abs/2206.15081} {arXiv:2206.15081 [astro-ph.CO]}
  \BibitemShut {NoStop}%
\bibitem [{\citenamefont {{Zhang}}\ \emph {et~al.}(2023)\citenamefont
  {{Zhang}}, \citenamefont {{Wang}}, \citenamefont {{Zhang}},\ and\
  \citenamefont {{Zhang}}}]{Zhang23}%
  \BibitemOpen
  \bibfield  {author} {\bibinfo {author} {\bibfnamefont {H.}~\bibnamefont
  {{Zhang}}}, \bibinfo {author} {\bibfnamefont {Y.-C.}\ \bibnamefont {{Wang}}},
  \bibinfo {author} {\bibfnamefont {T.-J.}\ \bibnamefont {{Zhang}}}, \ and\
  \bibinfo {author} {\bibfnamefont {T.}~\bibnamefont {{Zhang}}},\ }\href
  {\doibase 10.3847/1538-4365/accb92} {\bibfield  {journal} {\bibinfo
  {journal} {apjs}\ }\textbf {\bibinfo {volume} {266}},\ \bibinfo {eid} {27}
  (\bibinfo {year} {2023})},\ \Eprint {http://arxiv.org/abs/2304.03911}
  {arXiv:2304.03911 [astro-ph.CO]} \BibitemShut {NoStop}%
\bibitem [{\citenamefont {Rasmussen}\ and\ \citenamefont
  {Williams}(2006)}]{Rasmussen06}%
  \BibitemOpen
  \bibfield  {author} {\bibinfo {author} {\bibfnamefont {C.~E.}\ \bibnamefont
  {Rasmussen}}\ and\ \bibinfo {author} {\bibfnamefont {C.~K.~I.}\ \bibnamefont
  {Williams}},\ }\href@noop {} {\emph {\bibinfo {title} {Gaussian processes for
  machine learning.}}},\ Adaptive computation and machine learning\ (\bibinfo
  {publisher} {MIT Press},\ \bibinfo {year} {2006})\ pp.\ \bibinfo {pages}
  {I--XVIII, 1--248}\BibitemShut {NoStop}%
\end{thebibliography}%

\end{document}